\documentclass[aps,prl,twocolumn,showpacs,floatfix]{revtex4-1}
	
\usepackage{amsmath}
\usepackage{txfonts}
\usepackage{microtype}
\usepackage{graphicx}
\usepackage{color}
\usepackage{ulem}
\usepackage[english]{babel}

\begin{document}
\title{Under-the-tunneling-barrier recollisions in strong field ionization}

\author{Michael Klaiber}\email{klaiber@mpi-hd.mpg.de}
\affiliation{Max-Planck-Institut f\"ur Kernphysik, Saupfercheckweg 1, 69117 Heidelberg, Germany}
\author{Karen Z. Hatsagortsyan}\email{k.hatsagortsyan@mpi-hd.mpg.de}
\affiliation{Max-Planck-Institut f\"ur Kernphysik, Saupfercheckweg 1, 69117 Heidelberg, Germany}
\author{Christoph H. Keitel}
\affiliation{Max-Planck-Institut f\"ur Kernphysik, Saupfercheckweg 1, 69117 Heidelberg, Germany}

\date{\today}

\begin{abstract}

A new pathway of  strong laser field induced ionization
of an atom is identified which is based on recollisions under the tunneling barrier.
With an amended strong field approximation, the interference of
the direct and the under-the-barrier recolliding quantum orbits are
shown to induce a measurable shift of the peak of the photoelectron
momentum distribution. The scaling of the momentum shift is derived
relating the momentum shift to the tunneling delay time
according to the Wigner concept. This allows to extend the
Wigner concept for the quasistatic tunneling time delay into the
nonadiabatic domain. The obtained corrections to photoelectron momentum
distributions are also relevant for state-of-the-art accuracy of strong
field photoelectron spectrograms in general.

\end{abstract}

 \maketitle

Modern strong field photoelectron spectroscopy has achieved unprecedented momentum resolution of the order of $0.01$ atomics units (a.u.), see e.g. \cite{Blaga_2009,Dura_2013,Wolter_2015x}, due to advancement of the measurement technique with a reaction microscope \cite{Ullrich_2003}. Recently the attoclock technique has been developed  \cite{Eckle_2008a,Eckle_2008b} based on the strong field ionization of an atom in an elliptically polarized  laser field, which attempts to map the  photoelectron momentum at the detector into the time of the electron appearance in the continuum during strong field ionization. In this way the attoclock technique is assumed to extract information on the time-resolved dynamics of the electron released from the atomic bound state during strong field ionization, and  in particular, on the time-delay of the tunneling electron wave packet from the atom in a strong laser field \cite{Eckle_2008a,Eckle_2008b,Pfeiffer_2012,Landsman_2014o,Landsman_2014b,Camus_2017}. Furthermore, the interference structures in the high-resolution photoelectron momentum distribution (PMD),  created by the direct and recolliding trajectories, allow an interpretation  as time-resolved holographic imaging of atoms and molecules, which admits attosecond time- and \AA ngstr\"om spatial-resolution  \cite{Huismans_2011,Bian_2011,Marchenko_2011,Huismans_2012}. For a correct interpretation of  imaging results of the PMD based attoscience applications, one needs to understand theoretically all PMD features in details.

There are many theoretical approaches for the treatment of the tunneling delay time \cite{Landauer_1994,Sokolovski_2008,Landsman_2016}, leading to different solutions and to a debate on how to explain the photoelectron momentum distribution in attoclock experiments \cite{Landsman_2016,Torlina_2015,Ni_2016}. Although  all alternative definitions of the tunneling delay time are equally valid theoretical concepts, the Wigner concept \cite{Wigner_1955} is physically relevant  to the measurement of the photoelectron momentum distribution in the attoclock setup in the quasistatic regime, as proved in a recent experiment \cite{Camus_2017}. However the Wigner definition of the time delay via the derivative of the wave function phase, and its generalization for the strong field tunneling problem  \cite{Yakaboylu_2013,Yakaboylu_2014b,Torlina_2015,Kaushal_2015a,Kaushal_2015b} is applicable only in the quasistatic limit, i.e., when the laser induced barrier is (quasi-)static. Therefore, there is need for a generalization of the Wigner concept to the nonadiabatic regimes \cite{Yudin_2001b,Barth_2013a,Klaiber_2015} of the strong field ionization, which may explain the discrepancy between the theory and  the attoclock experiment at large Keldysh parameters \cite{Camus_2017}. 
\begin{figure}[b]
  \begin{center}
 \includegraphics[width=0.5\textwidth]{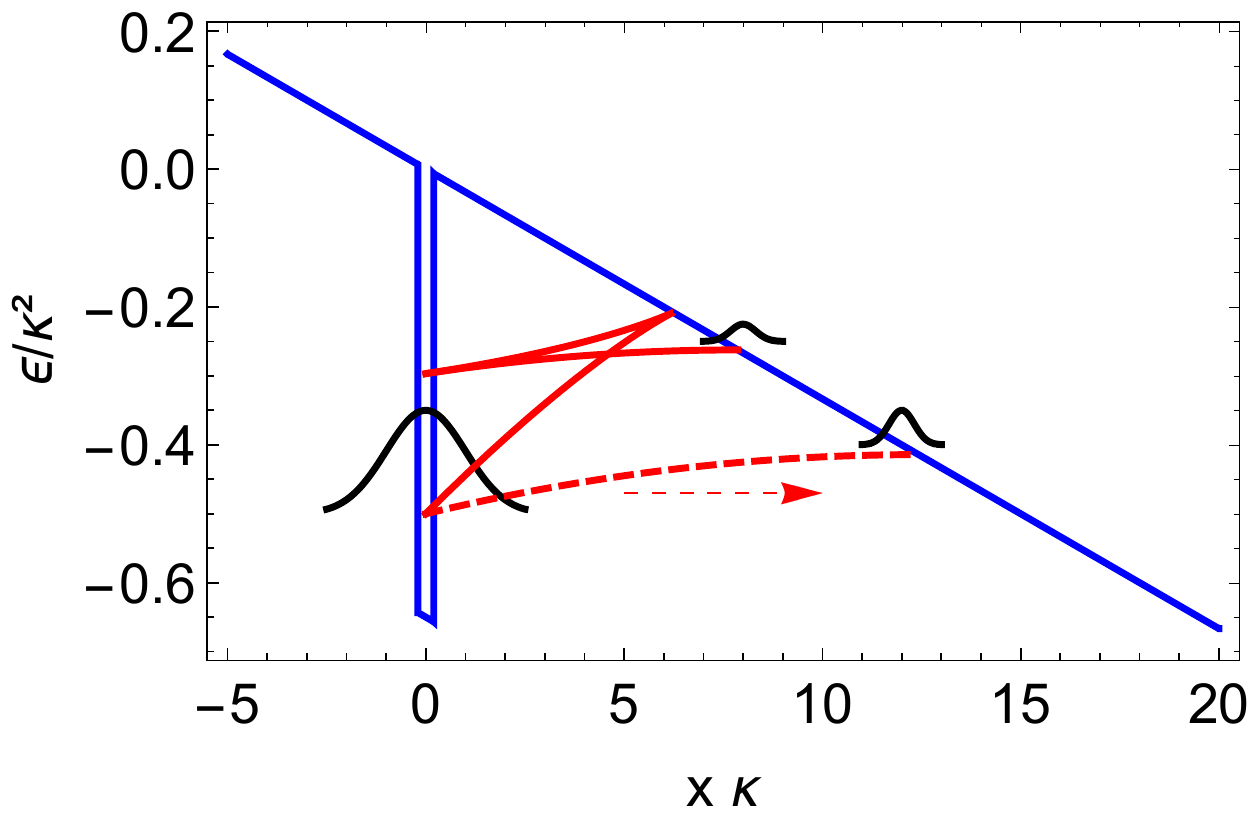}
           \caption{Schematic picture of laser-induced tunneling ionization: (dashed) the direct trajectory, and (solid) the under-the-barrier recolliding trajectory. The interference of the direct and the rescattered trajectories induces a shift of the peak of the photoelectron momentum distribution. The Keldysh-parameter is $\gamma=1$, featuring nonadiabatic tunneling, i.e., when the energy is not constant during tunneling.}
        \label{schema}
    \end{center}
  \end{figure}

The main workhorse for the theoretical treatment of the strong field ionization, the strong field approximation (SFA) \cite{Keldysh_1965,Faisal_1973,Reiss_1980}, in its common form does not provide a signature of the tunneling time in the asymptotic momentum distribution. The same is true for the Coulomb corrected SFA (CCSFA) \cite{Popruzhenko_2008a,Popruzhenko_2008b}, and the Analytic R-matrix (ARM) theory \cite{Torlina_2012,Torlina_2012b,Torlina_2013}, which include the Coulomb field of the atomic core for the continuum electron in the eikonal approximation [that is, in the Wentzel-Kramers-Brillouin (WKB) approximation combined with the perturbative accounting of the Coulomb field in the phase of the wave function]. To describe the Wigner tunneling time delay (emerging from the derivative of the phase of the wave function) within SFA, one needs to account for the phase of the wave function during the under-the barrier dynamics, which is vanishing in the leading order of WKB-approximation.  For the sake of intuitive understanding within an analytical treatment, this conceptual problem is most easily  addressed in the case of a short-range potential. It is well-known that the qualitative description of many strong field phenomena, such as  above-threshold ionization \cite{Becker_1989,Faria_2002}, high-order harmonic generation \cite{Becker_1990a,Becker_1994b}, or nonsequential double ionization \cite{Kopold_2000}, have been successfully given first in a simplified approach using a short-range potential.

In this Letter, we have modified the common SFA in the case of a short-range atomic potential, revealing and employing new quantum orbits for the ionizing electron, which describe rescattering of the electron at the atomic core during the under-the-barrier dynamics, see Fig.~\ref{schema}. We demonstrate that the interference of the direct and the under-the-barrier rescattering trajectories induces a phase shift of the wave function of the tunneling electron and a measurable shift of the peak of the momentum distribution. In the quasistatic regime the scaling of the momentum shift with respect to the laser and atom parameters is in accordance with the Wigner time delay theory, which allow us to interpret it accordingly. Moreover, the modified SFA provides a route for treating the Wigner time delay in nonadiabatic regimes of strong field ionization.

We consider the ionization of an atom in a laser field of linear polarization in the case of a short-range binding potential in the nonrelativistic regime. Ionization induced by a half cycle is considered, neglecting  interference effects from the ionization from neighbored half cycles. The Keldysh-parameter $\gamma=\kappa\omega/E_0$ is not restricted, with $\kappa=\sqrt{2I_p}$,  the ionization potential $I_p$,   the  laser field amplitude $E_0$ and frequency $\omega$, describing the tunneling, the multiphoton, as well as the transition regimes. The field strength parameter $f\equiv E_0/\kappa^3$ is assumed to be small to avoid over-the-barrier ionization, and atomic units are used throughout.
Having simplified the scenario to the basic physical process, we are able to
calculate the photoelectron momentum distribution $w(\textbf{p})=|M(\textbf{p})|^2$ analytically via a 2nd order SFA-amplitude \cite{Becker_2002}. For an improvement of the recollision treatment, the low frequency approximation  \cite{Cerkic_2009,Milosevic_2014} is employed, replacing the recollision matrix element in the Born approximation by the exact $T$-matrix:
\begin{eqnarray}
&  & M(\mathbf{p})= M_0(\mathbf{p})+M_1(\mathbf{p})=-i\int dt\langle\psi_\mathbf{p}(t)|H_i(t)|\phi(t)\rangle \label{mp1}\\
 &-&\int dt'\int^{t'}dt''\int d^3\mathbf{q}\langle\psi_\mathbf{p}(t')|T|\psi_\mathbf{q}(t')\rangle \langle\psi_\mathbf{q}(t'')|H_i(t'')|\phi(t'')\rangle, \nonumber
\end{eqnarray}
where $M_0,\,M_1$ are the direct and rescattering amplitudes, $|\psi_\mathbf{p}(t)\rangle=|\mathbf{p}+\mathbf{A}(t)\rangle\exp[S_\mathbf{p}(t)]/\sqrt{2\pi}^3$ is the Volkov-state in length gauge with the asymptotic momentum $\mathbf{p}$ and contracted action $S_\mathbf{p}(t)=\int^{\infty}_tds(\mathbf{p}+\mathbf{A}(s))^2/2$, $H_i(t)=-\mathbf{r}\cdot\mathbf{F}(t)$ the interaction Hamiltonian with the laser field induced force $\mathbf{F}(t)=E_0\mathbf{e}_x\cos(\omega t)$, $\partial_t \textbf{A}=\textbf{F}$, $\langle\mathbf{p}|H_i(t)|\phi\rangle$  the matrix element of the transition from the bound state into the continuum, $|\phi\rangle$  the initial bound state, and $\langle\mathbf{p}|T|\mathbf{q}\rangle$  the scattering $T$-matrix element.

First, we illustrate our theoretical approach in the 1D case, and further extend the discussion to 3D.
Thus, we begin considering the single active electron to be initially in its bound state in a 1D delta-potential $V(x)=-\kappa\delta(x)$, with the  wave function of the bound state
 $\langle x|\phi(t)\rangle=\sqrt{\kappa}\exp(-\kappa |x|+i\kappa^2/2t)$ \cite{Krajewska_2010}.
In 1D, $H_i=-x F(t)$ is $\langle p|H_i|\phi\rangle= 2\sqrt{2}i p F(t)/[\sqrt{\pi} (p^2+\kappa^2 )^2]$, and the exact scattering $T$-matrix is 
$\langle p|T|q\rangle=- (\kappa/2\pi) [\sqrt{p^2}/(\sqrt{p^2}-i\kappa)]$.
The momentum amplitude of Eq.~(\ref{mp1}) in 1D case ($d^3\mathbf{q}\rightarrow dq$) has two terms, 1D integral for the direct electron, and  3D integral for the rescattered electron. In the latter, rather than considering  the rescattering through the continuum excursion, which contribution is well investigated and takes place during at least two laser half cycles, we consider only  rescattering during the under-the-barrier motion which appears already in one laser half cycle.

For the physical interpretation of the recollision picture, we  firstly apply the simultaneous 3D saddle-point integration  analytically in the quasistatic case $\gamma\ll 1$, when the saddle point equations read:
\begin{eqnarray}
&& q_s =-E_0(t_r+t_i)/2, \nonumber\\
&& (p+E_0t_r)^2/2=(q_s+E_0t_r)^2/2, \\
&&(q_s+E_0t_i)^2/2=-I_p,\nonumber
\end{eqnarray}
which defines the intermediate momentum $q_s$ via the return condition of the trajectory, the recollision time  $t_r$ via the energy conservation at recollision, and the ionization time $t_i$ via the energy conservation at ionization. The saddle point equations yields the following physical solution $t_r=(-p+i\kappa)/E_0$ and $t_i=(-p+3i\kappa)/E_0$ (other solutions yield unphysical trajectories with increasing probabilities during propagation). Simplifying further for a moment with  $p=0$ and $\gamma\ll 1$, one obtains $t_i=3i\kappa/E_0$, $t_r=i\kappa/E_0$, and  $q_s+A(t_i)=i\kappa$, accordingly $q_s+A(t_r)=-i\kappa$ and $p+A(t_r)=i\kappa$. The latter provides the trajectory of the recolliding electron up to the recollision point:
\begin{eqnarray}
x(t)=i\kappa(t-t_i)+E_0(t-t_i)^2/2.
\end{eqnarray}
The trajectory starts at time $t_i$ at the atomic core $x(t_i)=0$, moves along the electric field through the barrier to the tunneling exit $x_e=I_p/E_0$, reaching it at $t=2i\kappa/E_0$. Afterwards the electron is reflected and turns around, tunnels  back to the core, where it recollides off the core $x=0$ at $t_r$, and again tunnels to the exit, leaving the barrier at $t_e=0$. In  Fig.~\ref{schema} the trajectory is visualized in the nonadiabatic regime at $\gamma=1$, with numerical solution of the sadle-point equations, showing the electron energy gain during ionization. 
\begin{figure}
  \begin{center}
  \includegraphics[width=0.23\textwidth]{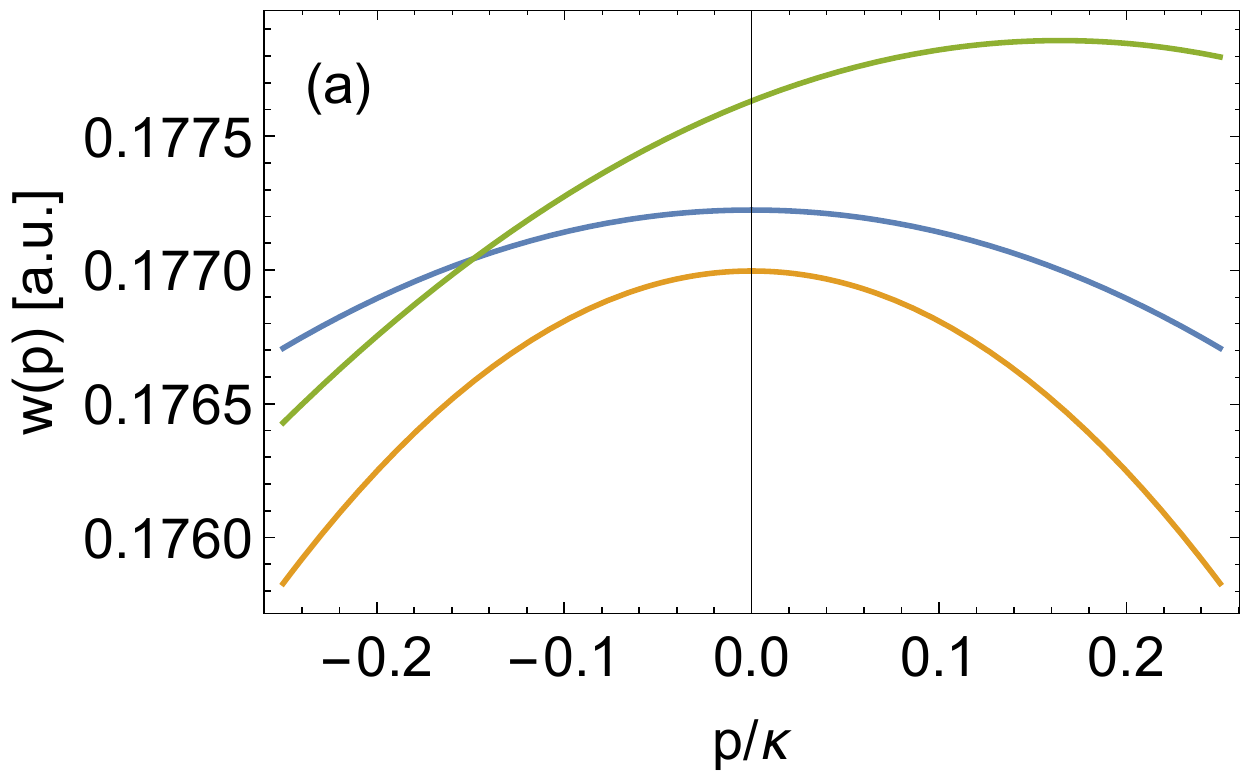}
\includegraphics[width=0.23\textwidth]{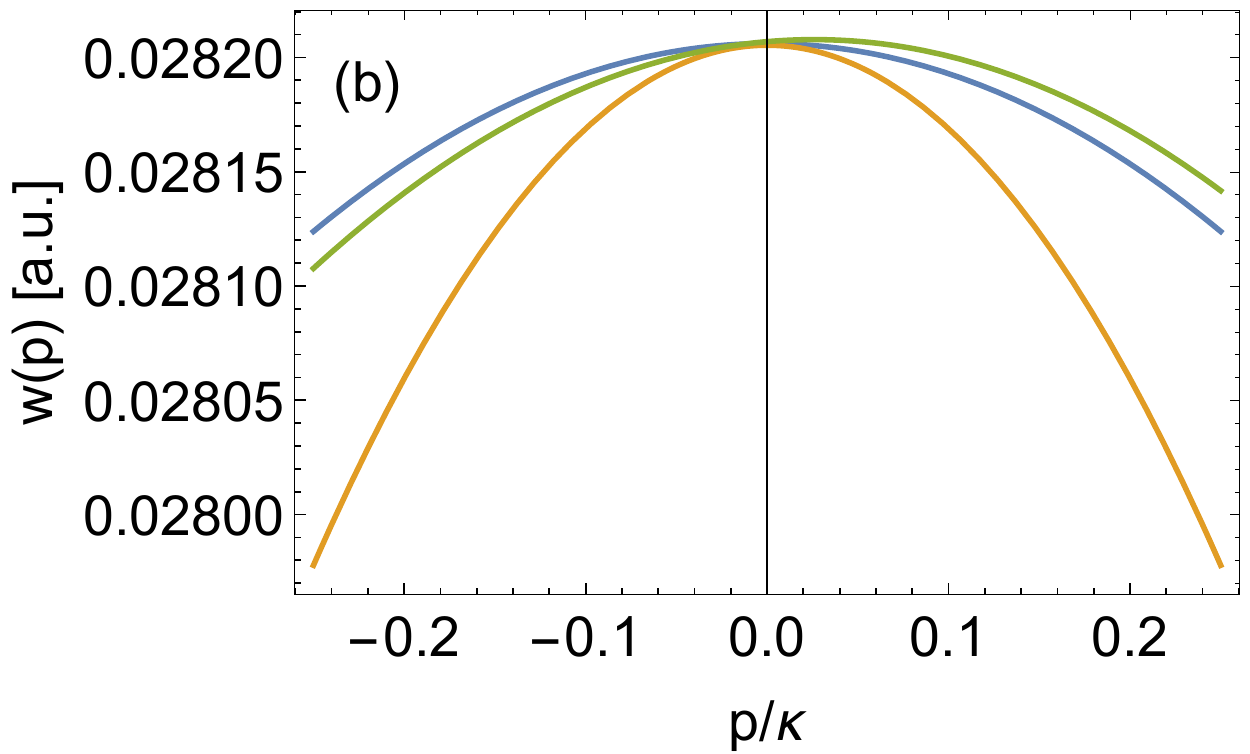}
           \caption{Photoelectron asymptotic momentum distribution  for $\gamma=0.2$ and $f=0.2$ in (a) 1D,  and (b) 3D:  (blue) via the direct amplitude, (brown) via the recolliding amplitude,  and (green) via including the interference of the direct and the recolliding trajectories. The brown curves are scaled by a factor of 434 in (a) and 33290 in (b), respectively.  In the case of a short-range potential $f=0.2$ corresponds to the below-threshold ionization, with the same tunneling exponent as $E_0\approx 1/22$ a.u. in the case of a Coulomb potential \cite{Suppl_material}.}
        \label{fig1}
    \end{center}
  \end{figure}

\begin{figure}
    \begin{center}
 \includegraphics[width=0.23\textwidth]{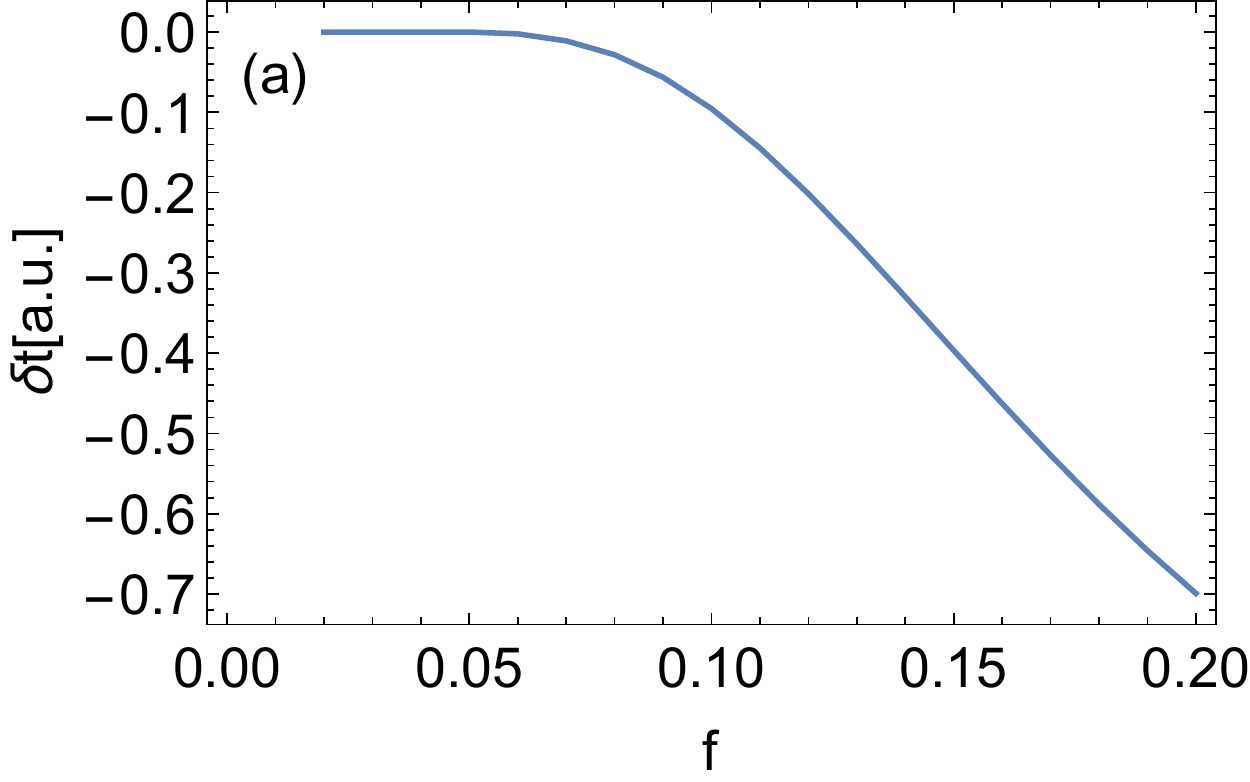}
 \includegraphics[width=0.23\textwidth]{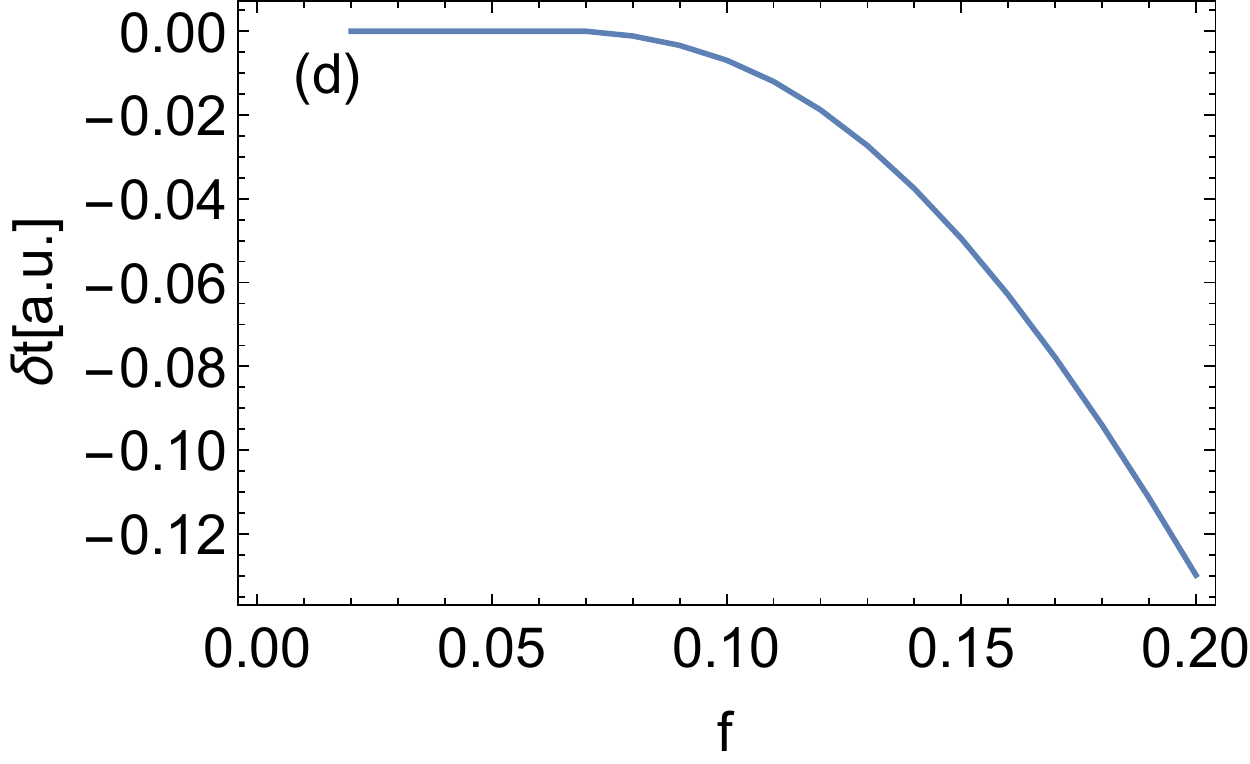}
 \includegraphics[width=0.23\textwidth]{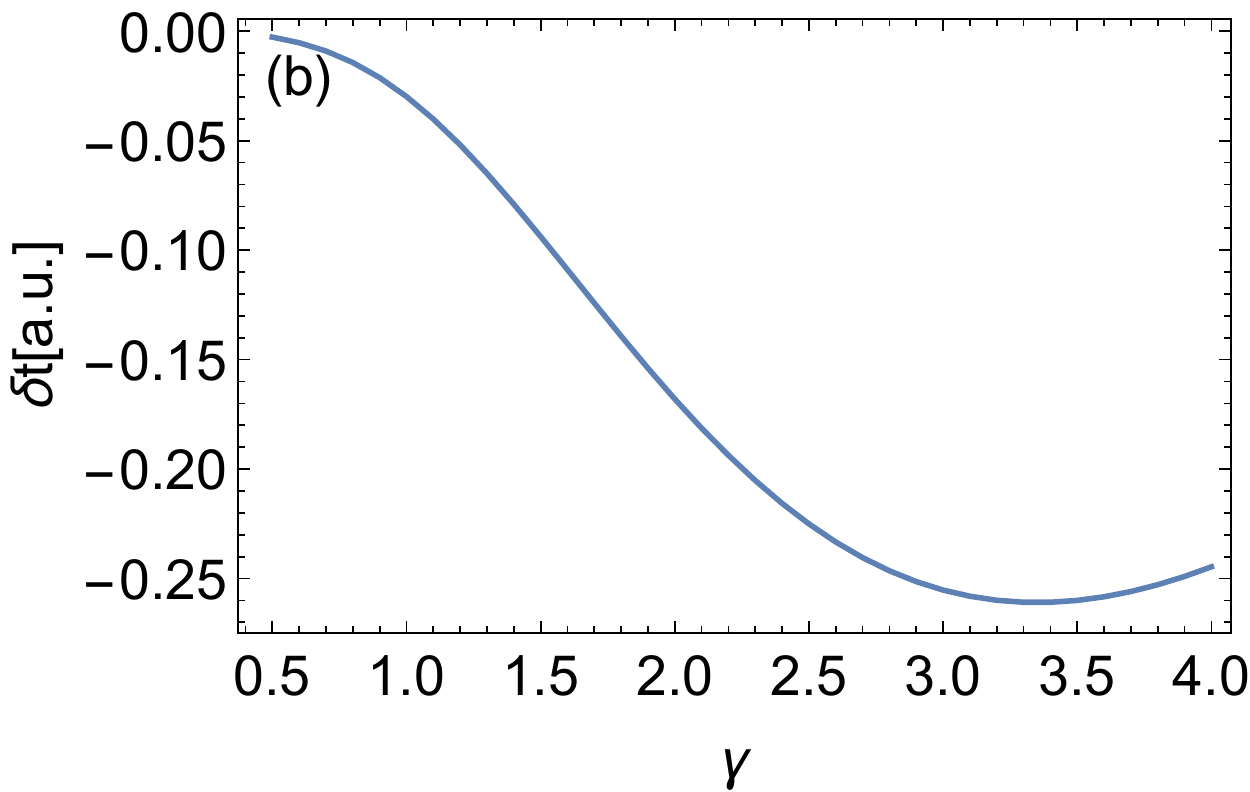}
 \includegraphics[width=0.23\textwidth]{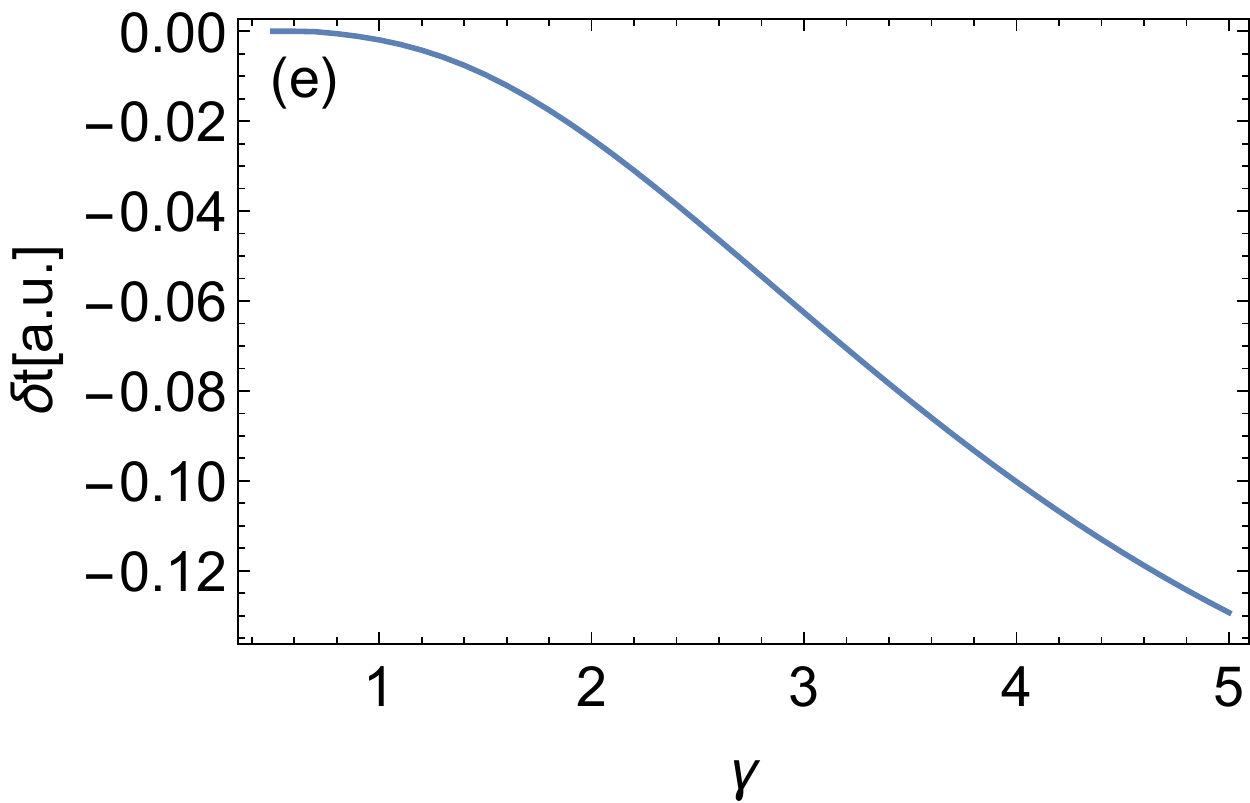}
 \includegraphics[width=0.23\textwidth]{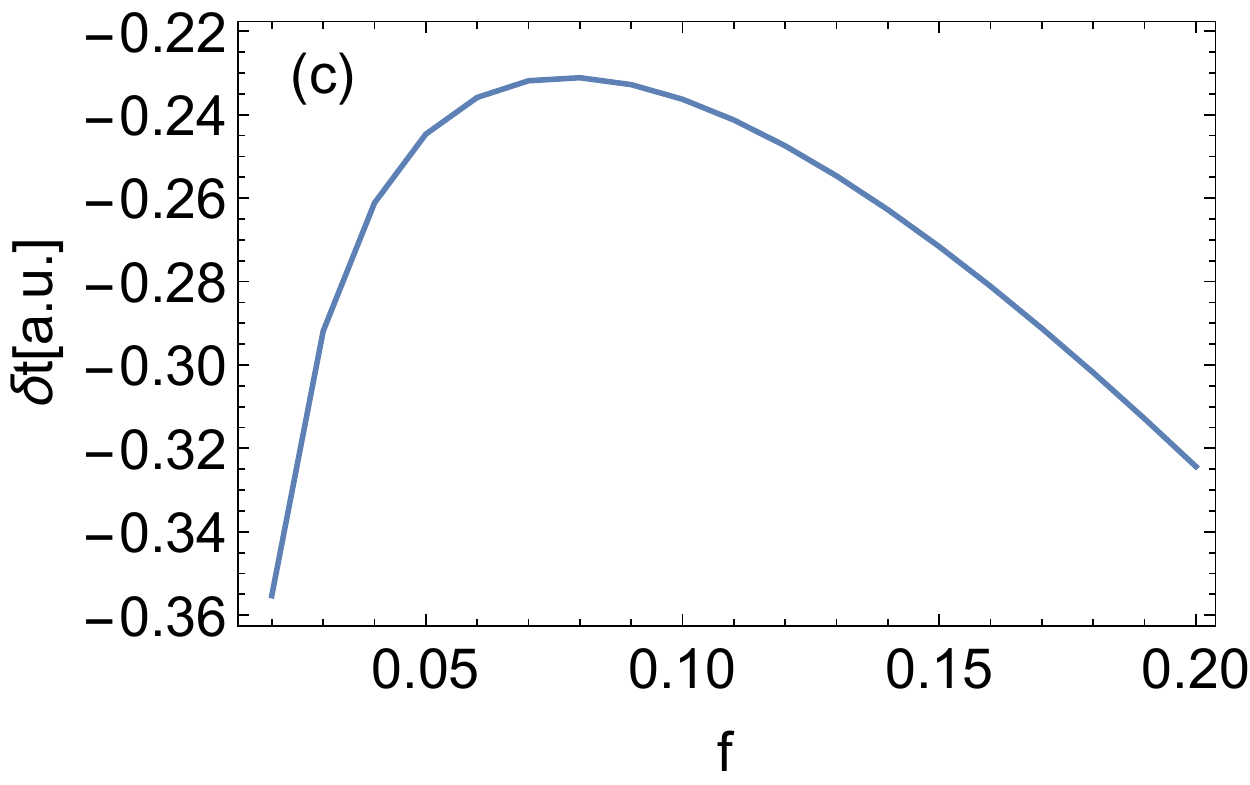}
 \includegraphics[width=0.23\textwidth]{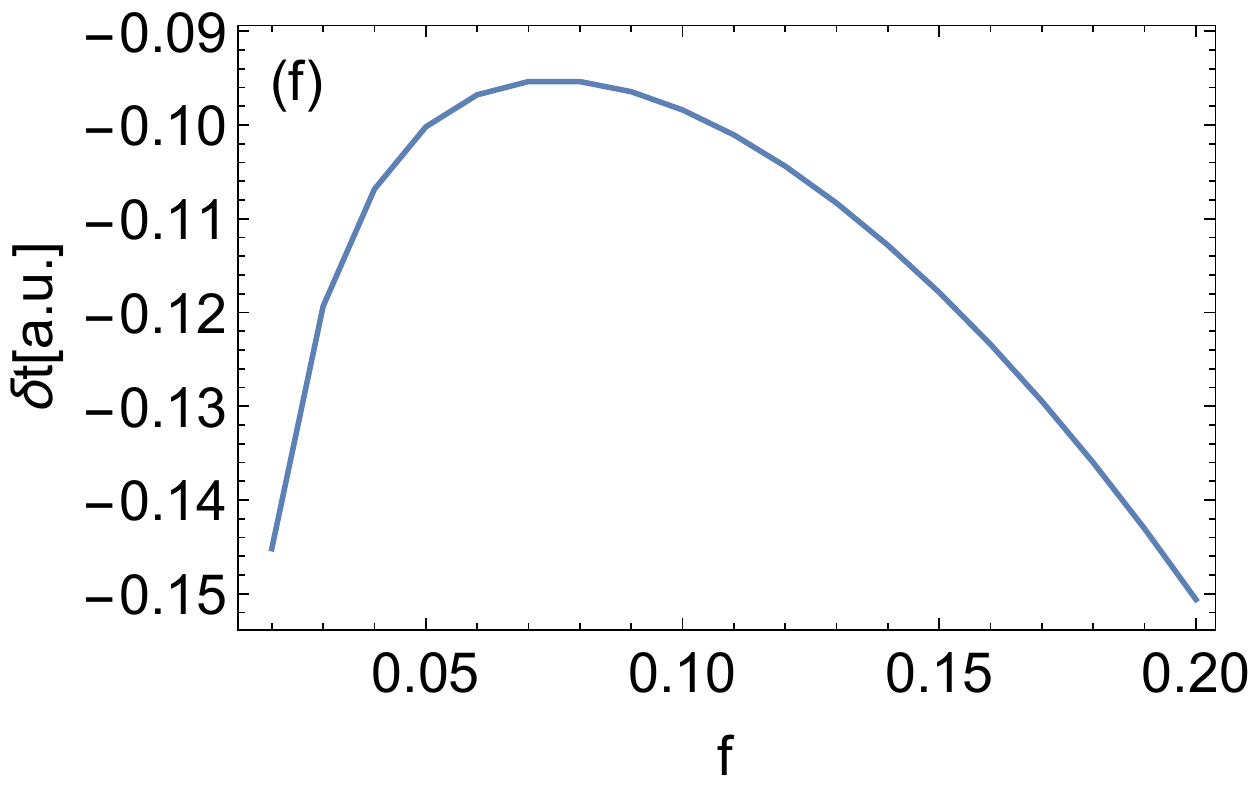} 
 \end{center}
           \caption{Tunneling time delay  versus the  field strength $f$ in the adiabatic regime of $\gamma=0.2$ [(a) 1D, (d) 3D],  versus the Keldysh-parameter $\gamma$ in the nonadiabatic regime for $f=0.05$ [(b) 1D, (e) 3D], and versus the  field strength  $f$ in the nonadiabatic regime for a fixed laser frequency of 
           $\omega=0.2$ a.u. [(c) 1D, (f) 3D]. }
        \label{fig2}
     \end{figure}
The accurate quantitative evaluation of the ionization amplitudes are carried out numerically. Both integrals in Eq.~(\ref{mp1}) are solved by exponentiation of the whole integrands:
\begin{eqnarray}
M(p)=-i \int dt \,e^{\zeta_0(t)}-\int dt'\int^{t'} dt''\int dqe^{\zeta_1(q,t',t'')}\nonumber
\end{eqnarray}
with $\zeta_0(t)=\ln(\langle\psi_p(t)|H_i(t)|\phi(t)\rangle)$ and $\zeta_1(q,t',t'')=\ln(\langle\psi_p(t')|T|\psi_q(t')\rangle\langle\psi_q(t'')|H_i(t'')|\phi(t'')\rangle)$
and applying the saddle-point method of integration. The functions $\zeta_0$ and $\zeta_1$ are expanded quadratically around the saddle points which are determined numerically, and the expanded function is  integrated analytically. The result is shown Fig.~\ref{fig1}.  Whereas the direct $|M_0(p)|^2$ and the recolliding $|M_1(p)|^2$ PMDs are peaked at zero  momentum, the coherent sum of the two distributions $|M_0(p)+M_1(p)|^2$ is slightly shifted towards positive momenta, i.e., the interference of the direct and the recolliding trajectories gives rise to  a momentum shift $\delta p$ of the PMD peak.

The behavior of the discussed momentum shift  in the quasistatic and the nonadiabatic regimes is illustrated in Figs.~\ref{fig2}(a) and~\ref{fig2}(b), respectively (the momentum shift $\delta p $ is equivalent to the time delay at the detector  $\delta t=-\delta p/E_0$; the momentum shift is positive which corresponds to the asymptotic negative time delay, see also \cite{Torlina_2015,Lein_2011}). In the quasistatic regime $\gamma\ll 1$ a significant tunneling time delay occurs when the field strength exceeds approximately 0.1 a.u. indicating that it is connected with near-threshold-tunneling. For smaller field strengths the recolliding path is strongly suppressed and do not affect the momentum distribution.
However, it also possible to have a significant momentum shift for relatively small field strength as long as the Keldysh-parameter is large, see Fig.~\ref{fig2}(b). The reason is that the electron gains  energy during the tunneling process and can enter in this way the near-threshold tunneling regime even for a small laser electric field strengths. Additionally, we display in Fig.~\ref{fig2}(c)   the tunneling time delay vs the field strength for a fixed laser frequency in the non-adiabatic regime,  corresponding to the typical experimental condition. When the frequency is fixed, a significant tunneling time delay occurs at large as well as at small field strengths, where the latter case again can be associated with a large Keldysh-parameter.

Let us estimate the scaling of momentum shift due to the interference of the direct and the under-the barrier trajectories in the quasi-static regime. The amplitudes of the direct electrons can be estimated as
$M_0\sim -i \delta t_i V_i\exp[-\kappa^3/(3E_0)]$ with the typical size of the volume element $\delta t_i\sim1/\sqrt{\partial^2_{t} S}\sim\sqrt{2\pi/(\kappa E_0)}$, and $V_i=\langle p|V|\phi\rangle=-  \kappa^{3/2}/\sqrt{2\pi}$, which yields
\begin{eqnarray}
M_0\sim\frac{ i}{\sqrt{f}}\exp\left(-\frac{1}{3f}\right).
\end{eqnarray}
Here, the equivalence of the ionization matrix element  $\langle p|H_i|\phi\rangle$ with $\langle p|V|\phi\rangle$ is used~\cite{Becker_2002}.

The amplitude of the rescattered  electrons can be estimated in the same way as $M_1\sim -\delta t_1 \delta t_2 \delta q  V_i \langle i\kappa|T|-i\kappa\rangle  \exp(-\kappa^3/E_0)/2$. The size of the volume element is $\delta t_1 \delta t_2 \delta q  \sim\sqrt{(2\pi)^3/(E_0\kappa^3)}$, which is estimated from the determinant of the matrix of the second order derivatives in the static regime, with  $\partial_{qq}S=-i(t_r-t_i)$, $\partial_{qt_1}S=-i E_0(t_r-t_i)/2$, $\partial_{qt_2}S=-i E_0(t_r-t_i)/2$, $\partial_{t_1t_1}S=i E_0^2(t_r+t_i)/2$, $\partial_{t_2t_1}S=0$, $\partial_{t_2t_2}S=-i E_0^2(t_r-t_i)/2$. Further, we estimate $\langle i\kappa|T|-i\kappa\rangle\sim\kappa/(2\pi\sqrt{ f})$, with the typical size of the recollision momentum $p+A(t_r)\sim i(\kappa+\sqrt{E_0/\kappa})$~\cite{Gribakin_1997}. Note that the recollision amplitude via the $T$-matrix is increased by a factor of $1/\sqrt{f}$, compared with the standard description in Born approximation, due to the singularity of the $T$-matrix at the  recollision energy of $-\kappa^2/2$.  Thus, the rescattering amplitude  is estimated to be: 
\begin{eqnarray}
M_1\sim-\frac{ 1}{f}\exp\left(-\frac{1}{f}\right).
\end{eqnarray}
Applying the quasi-static approximation, i.e., replacing  $E_0$  by the instantaneous electric field, and taking into account the time to momentum mapping, $E_0\rightarrow F(p)=E_0\sqrt{1-[\omega(-p+i \kappa)/E_0]^2}$ \cite{Delone_1991}, we obtain
\begin{eqnarray}
|M(p)|^2\sim\frac{\left|i\exp\left(-\frac{\kappa^3}{3F(p)}\right)-\frac{1}{\sqrt{f}}\exp\left(-\frac{\kappa^3}{F(p)}\right)\right|^2}{f}.
\end{eqnarray}
The latter has a maximum at $\delta p^{(1D)} \sim (M_1/M_0)\kappa\sim \exp[-2/(3f)]\kappa/\sqrt{f}$, demonstrating the PMD shift.
The amplitude of the recolliding electrons is smaller by a factor of $ \exp[-2/(3f)] $ due to the three times longer tunneling distance. In fact, an estimation of the tunneling amplitude via the WKB tunneling exponent $S=\int p dx$ along  the recolliding trajectory  yields $S=-\kappa^3/E_0$. Note that the replacement of the recollision matrix element in the Born approximation by the exact  $T$-matrix is necessary, because $p=i\kappa$   for the considered under-the-barrier recollision,  while the  Born approximation requires $p\gg \kappa$. Thus, the amplitude of the additional recolliding path is rather small, $M_0/M_1\approx 20$ as Fig. \ref{fig2}(a) shows. Nevertheless the momentum shift due to its interference with the direct trajectory is not negligible $\delta p\sim 0.1$.


 When  the under-the-barrier recollision scenario is applied in the 3D case \cite{Suppl_material}, the momentum distribution is qualitatively the same as in the 1D case, see Figs.~\ref{fig1}(b) and~\ref{fig2}(d),(e),(f), however the observed momentum shift is smaller, yielding $\delta p^{(3D)}\sim\exp[-2/(3f)]\kappa\sqrt{f}$. The reason is the spreading of the tunneling wave function under-the barrier in the 3D case of a zero-range potential, which reduces the recollision amplitude by a factor of $f$, and, consequently, decreases the momentum shift. In fact,  the intermediate momentum integration  yields in the 1D case a spreading factor of $\sqrt{2\pi/i\tau}$, whereas in 3D it is  $\sqrt{2\pi/i\tau}^3$ \cite{Ivanov_1996}, with the excursion time $\tau=t_r-t_i\sim{-2i\kappa/E_0}$. 

The described momentum shift due to  interference of the direct and the under-the barrier rescattered trajectories is closely related to the Wigner tunneling time delay. To demonstrate this, we recall the Wigner-formalism which accounts for the tunneling delay time during the laser-driven ionization process from a short-range atomic potential.
The time delay in the Wigner formalism is calculated as a derivative of the phase of the wave function. The continuum wave function in a slowly varying laser field (approximated by a constant electric field $E_0$), which has outgoing current and is matched with the bound state $\phi(x,p_y,p_z)$ at the matching coordinate under the barrier $x=x_m$~\cite{Suppl_material}, reads
\begin{eqnarray}
\psi(x,p_y,p_z)&=& {\cal T} \left[\text{Ai} ( \Xi   ) -i \text{Bi} (\Xi  )\right], \label{psi}
\label{Ai}
\end{eqnarray}
with the transition coefficient ${\cal T}  =\phi(x_m,p_y,p_z)/[\text{Ai}(\zeta)-i \text{Bi}(\zeta)]$, $\zeta\equiv  \sqrt[3]{2}( I_p-x_m E_0+p_y^2/2+p_z^2/2)/E_0^{2/3}$, and $\Xi\equiv \sqrt[3]{2} (I_p+p_y^2/2+p_z^2/2-E_0 x)/{E_0^{2/3}} )$.
The Wigner time delay is calculated as   \cite{Yakaboylu_2014b}
\begin{eqnarray}
\delta t= {\rm Re}\left\{ -i\frac{\partial}{\partial I_p}(\log[\psi(x,p_y,p_z)]-\log[\psi^{qc}(x,p_y,p_z)] \right\}\Big|_{x\rightarrow\infty}, 
\end{eqnarray}
where $\psi^{qc}(x,p_y,p_z)$ is the quasiclassical wave function, i.e., Eq.~(\ref{psi})  at the limit   $E_0\ll \kappa^3$, and the related momentum shift $\delta p=-E_0\delta t$ equals
\begin{eqnarray}
\delta p&=&\frac{\sqrt[3]{2} \sqrt[3]{f} \kappa }{\pi\left[ \text{Ai}\left(\frac{1}{2^{2/3} \tilde{f}(x_m)^{2/3}}\right)^2+  \text{Bi}\left(\frac{1}{2^{2/3} \tilde{f}(x_m)^{2/3}}\right)^2\right]} \sim \sqrt{\frac{\kappa^5}{E_0}}e^{-\frac{2 }{3\tilde{f}(0) }},\nonumber\\
\label{pW}
\end{eqnarray}
where $\tilde{f}(x)=E_0/(\kappa^2-2xE_0+p_y^2+p_z^2)^{3/2}$ is the reduced field, and the second equality is valid at  small field asymptotics. The Wigner  momentum shift of Eq.~(\ref{pW}) depends on the transversal momenta $p_y$  and $p_z$. Assuming that this momentum shift is measured by a detector on the $x$-axis at $y=0$ and  $z=0$, and that all transversal momenta contribute to the wavefunction at this position we weight the momentum shift with the bound state probability density  $|\phi(x=x_m,p_y,p_z)|^2\approx (2\pi/\kappa)\exp(-p_y^2/\kappa^2-p_z^2/\kappa^2)$, which yields an effective momentum shift of $\delta p^{(3D)}=f\delta p|_{p_y=p_z=0}$  again reduced by a factor of $f$ compared to the 1D-case. From the latter one can deduce that the derived 3D Wigner momentum shift coincides with the momentum shift due to interference of the direct and the under-the-barrier recolliding trajectories discussed above. 

 Up to now we have discussed the case of a linearly polarized laser field. However, the experimental observation of the discussed momentum shift will require the attoclock setup (elliptically polarized laser field close to circular) to avoid masking the effect by the low-energy structures \cite{Blaga_2009,Dura_2013,Wolter_2015x}. To see if the effect is modified in the attoclock setup, we   have calculated  the shift of the momentum distribution peak in the  case of elliptical polarization \cite{Suppl_material}.  We analyzed separately the role of the under-the-barrier recollisions, and the recollisions in the continuum. There is no shift in the momentum distribution due to interference of the direct and rescattered in the continuum trajectories, however, there is a shift due to interference of the direct and the under-the-barrier rescattered  trajectories. The magnitude of the shift fits to the case of linear polarization.  This is intuitively explainable, because there is no significant variation of the tunneling barrier and of the under-the-barrier recollisions in the quasistatic regime as in the linear as well as in the circular polarization cases.

In the present discussion the effect of the Coulomb field of the atomic core is neglected in the description of the ionization process. With the Coulomb field taken into account the tunneling process in the static regime takes place along one of the parabolic coordinates. In the transverse direction to the tunneling coordinate the electron dynamics is confined by a channel which would suppress the spreading and increase the recollision probability. We may therefore expect that in a realistic situation with the Coulomb field of the atomic core in action, the under-the-barrier rescattering process would be more similar to the 1D case with a short-range potential, than to the 3D short-range potential case, and the discussed momentum shift will be significant for strong field ionization in the near threshold regime.

Concluding, we have found a new type of rescattering trajectories during the under-the barrier dynamics in strong field tunneling ionization, and demonstrate that interference of the direct and the under-the-barrier recolliding trajectories induces a shift of the peak of the photoelectron momentum distribution. We advocate that the observed shift coincides with the momentum shift due to the Wigner tunneling time delay. It is of the order of 0.1 a.u. which translated into time delay corresponds to tens of attoseconds in the near-threshold regime and measurable with present experimental accuracy \cite{Blaga_2009,Dura_2013,Wolter_2015x,Camus_2017}.

MK acknowledges useful discussions with John Briggs.

\pagebreak

\begin{center}
\large{{\bf Supplementary Materials}}
\end{center}

\section{Recollisions in the case of a linearly polarized laser field with a 3D short-range potential }

Let us apply the under-the-barrier recollision scenario  in the 3D case.
The active electron in the free atomic system is in the ground state $\langle \mathbf{r}|\phi(t)\rangle=\sqrt{\kappa/(2\pi r^2)}\exp(-\kappa r+i\kappa^2/2t)$ of a 3D short-range potential $V(\mathbf{r})=\frac{2\pi}{\kappa}\delta(\mathbf{r})\partial_r r$, with $r=\sqrt{x^2+y^2+z^2}$ [44].
The ionization amplitude of Eq.~(1) of the main text includes the transition amplitude $\langle\mathbf{p}|H_i(t)|\phi\rangle=(2i/\pi)\sqrt{\kappa } \mathbf{F}(t)\cdot\mathbf{p}/ (\kappa ^2+p^2)^2$, and the scattering $T$-matrix element  $\langle\mathbf{p}|T|\mathbf{q}\rangle=-i/[4\pi^2(p-i \kappa)]$,
with $p=\sqrt{p_x^2+p_y^2+p_z^2}$ .
The integrals in Eq.~(1) are calculated with the help of a numerical saddle point approximation in analogy to the 1D-case. The results for the  photoelectron momentum distribution are shown in Figs.~(1b) and~(2d),(2e),(2f) of the main text. The momentum distribution is qualitatively the same as in the 1D case, however the observed momentum shift is smaller, yielding $\delta p^{(3D)}\sim\exp[-2/(3f)]\kappa\sqrt{f}$. The reason is the spreading of the tunneling wave function under-the barrier in the 3D case of a zero-range potential, which reduces the recollision amplitude by a factor of $f$, and, consequently, decreases the momentum shift. In fact,  the intermediate momentum integration 
yields in the 1D case a spreading factor of $\sqrt{2\pi/i\tau}$, whereas in 3D it is  $\sqrt{2\pi/i\tau}^3$ [48], with the excursion time $\tau=t_r-t_i\sim{-2i\kappa/E_0}$.

 \section{The under-the-barrier recollisions in an elliptically polarized laser field}

Here we present result of calculations of photoelectron momentum distribution for the ionization in an elliptically polarized laser field (attoclock).  We analyze the role of the under-the-barrier recollisions for the  shift of the peak of the momentum distribution, comparing the ionization spectrum from direct electrons with an ionization spectrum that also includes the contribution from the recolliding electrons.  Two type of recollisions are investigated: the recollisions in the continuum, and the  under-the-barrier recollisions.

In  the driving laser field is described by its vector potential 
\begin{eqnarray}
\mathbf{A}(t)=(E_0/\omega) f_e(t)[\cos(\omega t)\mathbf{e}_x-e \sin(\omega t)\mathbf{e}_y],
\end{eqnarray}
with the Gaussian envelope $f_e(t)=\exp(-(t/\tau)^2/2)$. The  applied parameters are: the laser field strength  $E_0=0.2$ a.u., the laser angular frequency $\omega=0.04$ a.u., the ellipticity $e=0.8$ and the pulse length $\tau=5/\omega$. The  potential of the atomic core is modeled by a 3D-short-range potential 
$V=2\pi/\kappa\delta(\mathbf{r})\partial_r r$ with $\kappa=1$, and a bound 
state $|\phi(t)\rangle$ also given in the main text. These parameters yield  a Keldysh-parameter of $\gamma=0.2$ and a field strength parameter of $f=0.2$. From this it follows that we are considering in the following the quasi-static near-threshold tunnel ionization regime.  
The field strength of $E_0=0.2$ a.u. in a short-range potential can be translated into the Coulomb case via the relation 
\begin{eqnarray}
-i\int^{x_e}_{x_i}dx p_C(x)=\frac{\kappa^3}{3f}
\end{eqnarray}
with the momentum in the Coulomb potential $p_c(x)=\sqrt{2(-1/2+E_0 x+1/x)}$, $x_i\approx2$ and $x_e\approx1/2/E_0$ yielding $E_0\approx 1/22$ a.u. The relation means that the effective potential hill in the Coulomb and the short-range potential case give the same tunneling exponent 
when the field strength are 1/22 a.u. or 0.2 a.u. respectively.

\subsection{The spectrum due to direct electrons}

\begin{figure}
    \begin{center}
      \includegraphics[width=0.4\textwidth]{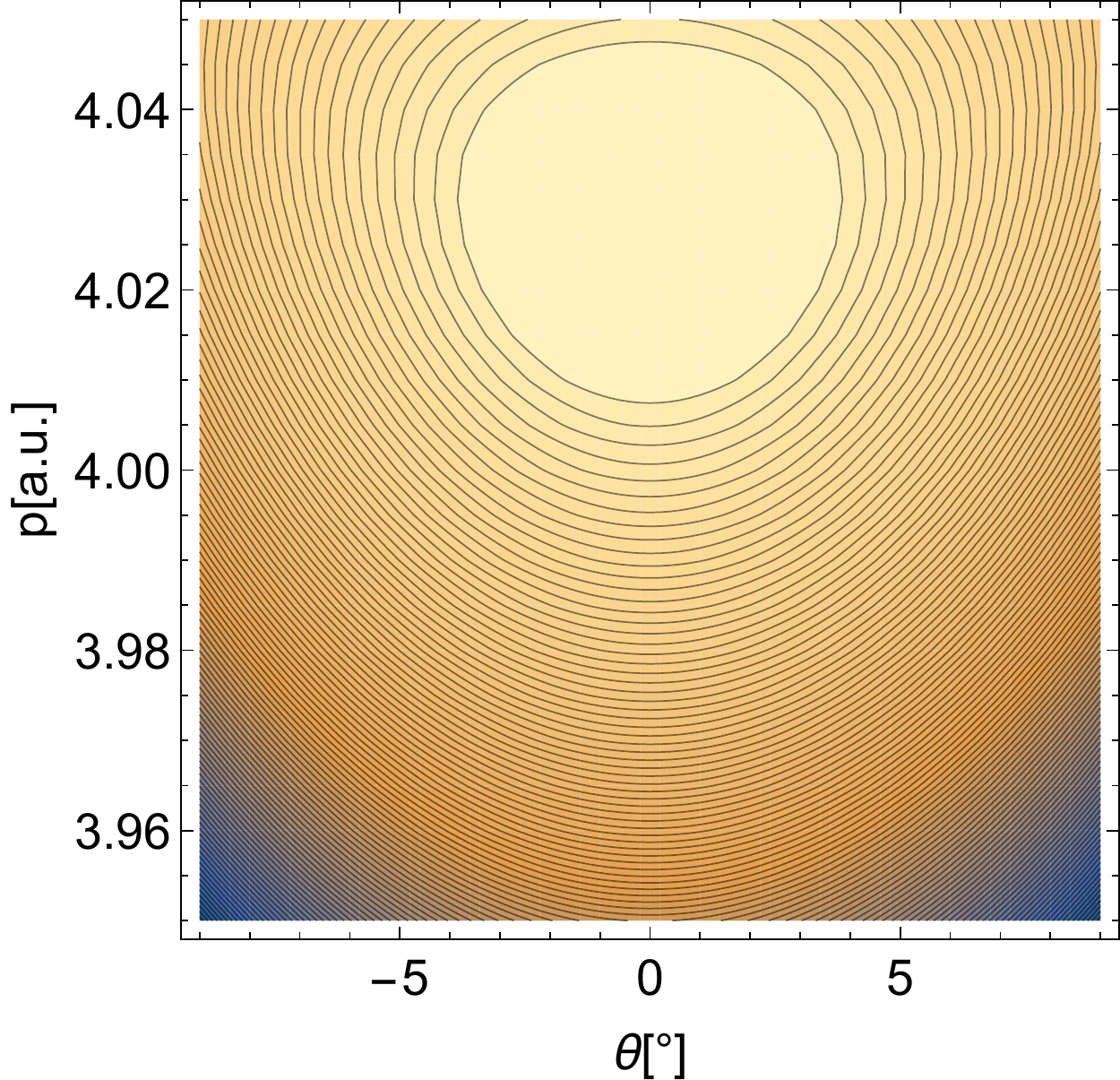} 
            \caption{Momentum distribution of the direct electrons vs the final momentum $p=\sqrt{p_x^2+p_y^2}$ and the emission angle $\theta=\arctan(p_x/p_y)$.}
        \label{md}
    \end{center}
  \end{figure}

The momentum distribution $w(\mathbf{p})=|M_0(\mathbf{p})|^2$ of the direct electrons, i.e. electrons that do not interact with the atomic core a second time after ionization can be given via the SFA-amplitude, see Eq.~(1) of the paper, where the vector potential of the linearly polarized field should be replaced by the elliptical one. After integrating the expression with the saddle point method the spectrum peaks at approximately $p=eE_0/\omega+e\gamma\kappa/6$ and $\theta=0^{\circ}$, see Fig.~\ref{md}, which is consistent with the simple-man-model including non-adiabatic corrections in the final momentum.

\subsection{The spectrum due to the interference of direct and the under-the-barrier recolliding electrons}

\begin{figure}
    \begin{center}
      \includegraphics[width=0.4\textwidth]{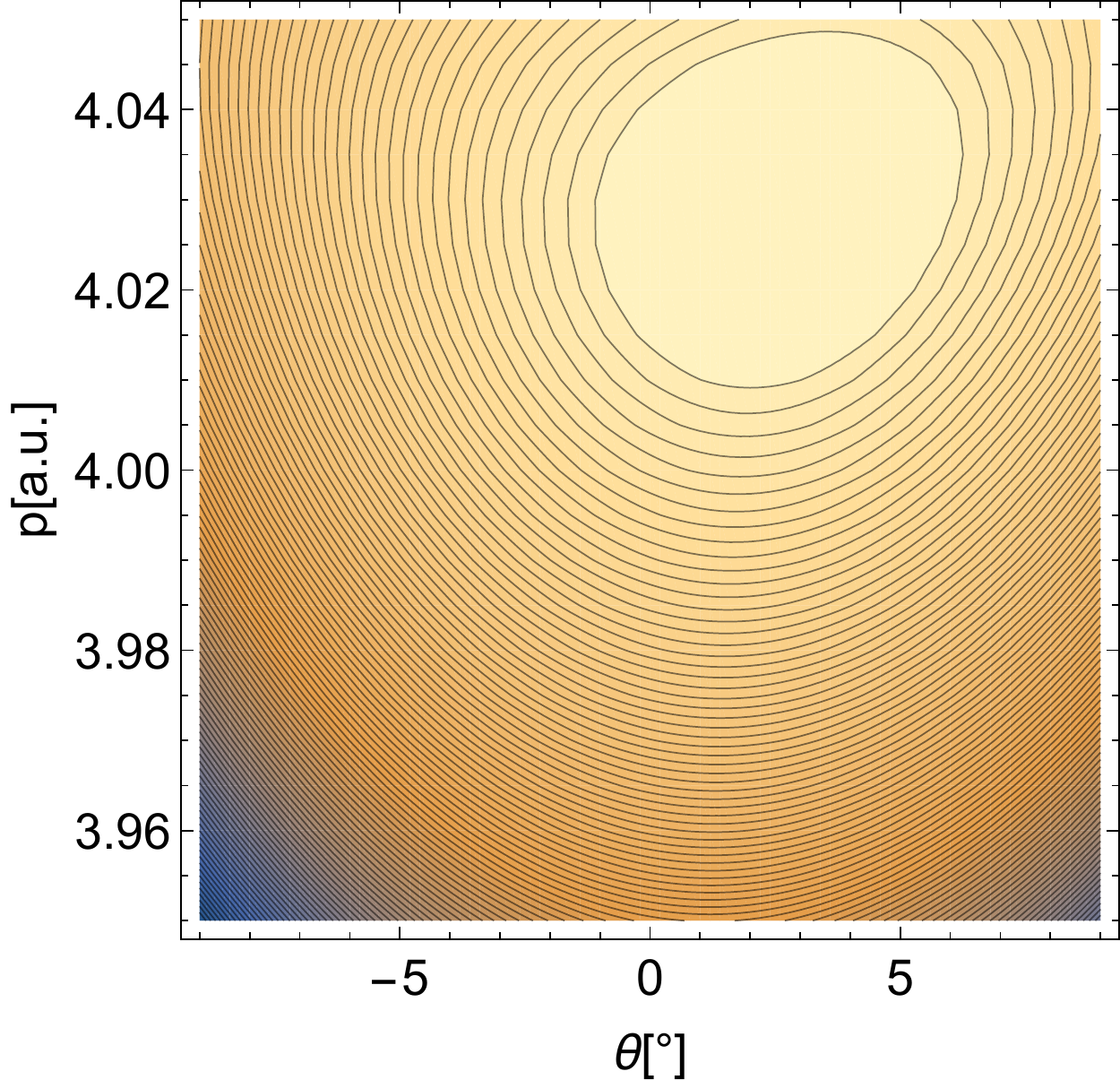} 
            \caption{ Momentum distribution of the direct and the recolliding electrons vs the final momentum $p=\sqrt{p_x^2+p_y^2}$ and the emission angle $\theta=\arctan(p_x/p_y)$.}
        \label{mu}
    \end{center}
  \end{figure}

The momentum distribution $w(\mathbf{p})=|M_0(\mathbf{p})+M_1(\mathbf{p})|^2$ represents the interference of direct and the under-the-barrier recolliding electrons, which is calculated via Eq.~(1) from the main text. In Fig.~\ref{mu} the spectrum is displayed and one can see a shift of the peak to an angle of approximately $2^{\circ}$.
\begin{figure}
    \begin{center}
      \includegraphics[width=0.4\textwidth]{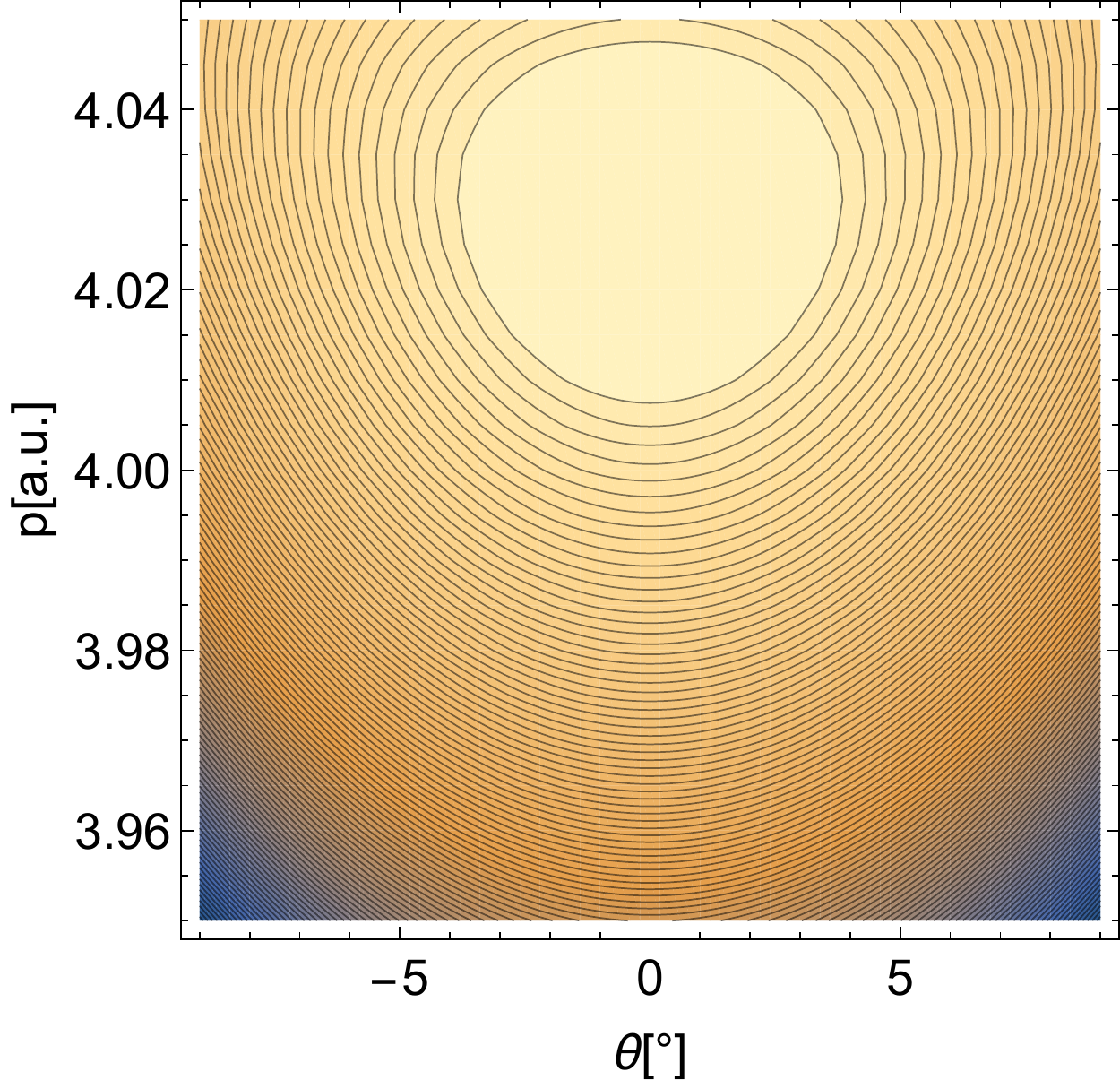} 
            \caption{ Momentum distribution of the direct and the recolliding electrons in the continuum vs the final momentum $p=\sqrt{p_x^2+p_y^2}$ and the emission angle $\theta=\arctan(p_x/p_y)$.}
        \label{mr}
    \end{center}
  \end{figure}

In the calculation of intermediate momentum integral in $M_1$ the saddle point approximation is applied which yields a preexponential term $\sim \sqrt{-2i\pi/(t_1-t_2)}^3$ with the ionization and recollision time $t_2$ and $t_1$, respectively. This term can now be associated with the spreading of
the wave-packet during tunneling. Since we want to drop this physical effect especially perpendicular to the tunneling direction to mimic a Coulomb potential, we replace the second order derivatives of the function in the exponent of Eq.~(1) in the main text by the second order derivatives of
the bound state:
\begin{eqnarray}
i(t_1-t_2) \rightarrow \partial_{q_yq_y}\ln(\phi(\mathbf{q}))|_{\mathbf{q}=0}=-\frac{2}{\kappa^2}\nonumber\\
i(t_1-t_2)  \rightarrow \partial_{q_zq_z}\ln(\phi(\mathbf{q}))|_{\mathbf{q}=0}=-\frac{2}{\kappa^2}
\end{eqnarray}
With this replacement the preexponential factor that is used in the calculation reads $\sim \sqrt{(-2i\pi)^3/(t_1-t_2)/(2/\kappa^2)^2}$. 

With this the determined emission angle of two degrees corresponds to an asymptotic tunneling time delay of $\theta/\omega\sim \pi/(90\omega)\approx 0.7 $ a.u. and is consistent with the 1D-result in the static regime from the main text. This is not surprising since the applied parameters are in the quasi-static regime of small Keldysh-parameters where during tunneling only the instantaneous value of the laser field strength enters the calculation and the field rotation can be neglected.

\subsection{Correction due to the recolliding electrons in the continuum }

Let us also estimate the contribution of electrons that recollide with the atomic core after an excursion in the continuum. Due to the elliptically polarized laser field the electron gain a momentum $e E_0/\omega$ in the $y$-direction. This momentum prevents the ionized  electrons of returning to the core. The only possibility that they can still come back is that they are ionized with a initial momentum $p_{y,i}$ at the exit that compensates this momentum shift due to the laser field. The ionization probability is then given by the well-known tunneling exponent
\begin{eqnarray}
\Gamma\sim\exp\left[-\frac{2(\kappa^2+p_{y,i}^2)^{3/2}}{E_0}\right]
\end{eqnarray}
which yields for $p_{y,i}\sim-e E_0/\omega$ a negligible small value of $10^{-100}$. We can therefore conclude that continuum recolliding electrons are strongly suppressed compared to direct electrons that are ionized with $p_{z,i}\sim 0$ and do not affect the momentum distribution, which Fig.~\ref{mr} illustrate.

\subsection{The role of the ionization p-state}

In the attoclock setup the ionization of the atom usually takes place from a p-state.  In the quasi-static regime the ionization from $p$-state happens from the orbital with $m=0$, where $m$ is magnetic quantum number with respect to the electric field strength. This orbital's wavefunction  behaves $\sim p_E/(i\kappa)\psi_{s}$, where $p_E$ is the momentum at the moment of ionization in direction of the laser electric field and $\psi_s$ is the wavefunction of an $s$-state. Therefore, at the moment of ionization, where $p_E\sim i \kappa$ the active $p$-electron wave function behavior is similar to that of an $s$-electron and there will be no deviation of the signatures  in the momentum distribution.

 \section{Recollision length in the low-frequency approximation}

The physical picture of the recollision in the Born-approximation is the following: the electron approaches the core from the left via a wave function $\exp(i p x)$ with $p=-i\kappa$, at the core ($x=0$) it is reflected, and leaves the core with the wave function $f_B\exp(i p' x)$ with the scattering amplitude in the Born-approximation $f_B=1$ and the outgoing momentum $p=i\kappa$.

In the improved description via the low-frequency approximation the scattering amplitude is enhanced. The scattering amplitude reads $f=-f_B i\kappa/(p-i\kappa)$ and with the typical momentum $p=i(\kappa+\sqrt{E_0/\kappa})$ it follows that the enhancement factor between the low-frequency and the Born-approximation is $\sqrt{\kappa^3/E_0}$. These results can now be interpreted as reflection at $x=0$ with a reflection coefficient larger than 1. An alternative physical interpretation can be given that the reflexion coefficient is still 1, but the reflection happens not at the core, but at an $x_m>0$, more precisely at an $x_m$ that is defined by the condition $\sqrt{\kappa^3/E_0}\exp(-\kappa x_m)=\exp(\kappa x_m)$, i.e. at $x_m=\ln(\sqrt{\kappa^3/E_0})/(2\kappa)\sim -\ln(f)/(4\kappa)$. 

From this argumentation one may expect that the  coordinate $x_m$ defines also the point where the Wigner trajectory starts. Therefore, the initial point of the Wigner-trajectory employed in the text of the paper, we use the coordinate $x_m$ derived above.

\bibliography{strong_fields_bibliography}

\begin{thebibliography}{49}
\expandafter\ifx\csname natexlab\endcsname\relax\def\natexlab#1{#1}\fi
\expandafter\ifx\csname bibnamefont\endcsname\relax
  \def\bibnamefont#1{#1}\fi
\expandafter\ifx\csname bibfnamefont\endcsname\relax
  \def\bibfnamefont#1{#1}\fi
\expandafter\ifx\csname citenamefont\endcsname\relax
  \def\citenamefont#1{#1}\fi
\expandafter\ifx\csname url\endcsname\relax
  \def\url#1{\texttt{#1}}\fi
\expandafter\ifx\csname urlprefix\endcsname\relax\def\urlprefix{URL }\fi
\providecommand{\bibinfo}[2]{#2}
\providecommand{\eprint}[2][]{\url{#2}}

\bibitem[{\citenamefont{Blaga et~al.}(2009)\citenamefont{Blaga, Catoire,
  Colosimo, Paulus, Muller, Agostini, and DiMauro}}]{Blaga_2009}
\bibinfo{author}{\bibfnamefont{C.~I.} \bibnamefont{Blaga}},
  \bibinfo{author}{\bibfnamefont{F.}~\bibnamefont{Catoire}},
  \bibinfo{author}{\bibfnamefont{P.}~\bibnamefont{Colosimo}},
  \bibinfo{author}{\bibfnamefont{G.~G.} \bibnamefont{Paulus}},
  \bibinfo{author}{\bibfnamefont{H.~G.} \bibnamefont{Muller}},
  \bibinfo{author}{\bibfnamefont{P.}~\bibnamefont{Agostini}}, \bibnamefont{and}
  \bibinfo{author}{\bibfnamefont{L.~F.} \bibnamefont{DiMauro}},
  \bibinfo{journal}{Nat. Phys.} \textbf{\bibinfo{volume}{5}},
  \bibinfo{pages}{1745} (\bibinfo{year}{2009}).

\bibitem[{\citenamefont{Dura et~al.}(2013)\citenamefont{Dura, Camus, Thai,
  Britz, Hemmer, Baudisch, Senftleben, Schr{\"{o}}ter, Ullrich, Moshammer
  et~al.}}]{Dura_2013}
\bibinfo{author}{\bibfnamefont{J.}~\bibnamefont{Dura}},
  \bibinfo{author}{\bibfnamefont{N.}~\bibnamefont{Camus}},
  \bibinfo{author}{\bibfnamefont{A.}~\bibnamefont{Thai}},
  \bibinfo{author}{\bibfnamefont{A.}~\bibnamefont{Britz}},
  \bibinfo{author}{\bibfnamefont{M.}~\bibnamefont{Hemmer}},
  \bibinfo{author}{\bibfnamefont{M.}~\bibnamefont{Baudisch}},
  \bibinfo{author}{\bibfnamefont{A.}~\bibnamefont{Senftleben}},
  \bibinfo{author}{\bibfnamefont{C.~D.} \bibnamefont{Schr{\"{o}}ter}},
  \bibinfo{author}{\bibfnamefont{J.}~\bibnamefont{Ullrich}},
  \bibinfo{author}{\bibfnamefont{R.}~\bibnamefont{Moshammer}},
  \bibnamefont{et~al.}, \bibinfo{journal}{Scientific Reports}
  \textbf{\bibinfo{volume}{3}}, \bibinfo{pages}{2675} (\bibinfo{year}{2013}).

\bibitem[{\citenamefont{Wolter et~al.}(2015)\citenamefont{Wolter, Pullen,
  Baudisch, Sclafani, Hemmer, Senftleben, Schr\"oter, Ullrich, Moshammer, and
  Biegert}}]{Wolter_2015x}
\bibinfo{author}{\bibfnamefont{B.}~\bibnamefont{Wolter}},
  \bibinfo{author}{\bibfnamefont{M.~G.} \bibnamefont{Pullen}},
  \bibinfo{author}{\bibfnamefont{M.}~\bibnamefont{Baudisch}},
  \bibinfo{author}{\bibfnamefont{M.}~\bibnamefont{Sclafani}},
  \bibinfo{author}{\bibfnamefont{M.}~\bibnamefont{Hemmer}},
  \bibinfo{author}{\bibfnamefont{A.}~\bibnamefont{Senftleben}},
  \bibinfo{author}{\bibfnamefont{C.~D.} \bibnamefont{Schr\"oter}},
  \bibinfo{author}{\bibfnamefont{J.}~\bibnamefont{Ullrich}},
  \bibinfo{author}{\bibfnamefont{R.}~\bibnamefont{Moshammer}},
  \bibnamefont{and} \bibinfo{author}{\bibfnamefont{J.}~\bibnamefont{Biegert}},
  \bibinfo{journal}{Phys. Rev. X} \textbf{\bibinfo{volume}{5}},
  \bibinfo{pages}{021034} (\bibinfo{year}{2015}).

\bibitem[{\citenamefont{Ullrich et~al.}(2003)\citenamefont{Ullrich, Moshammer,
  Dorn, D\"orner, Schmidt, and Schmidt-B\"ocking}}]{Ullrich_2003}
\bibinfo{author}{\bibfnamefont{J.}~\bibnamefont{Ullrich}},
  \bibinfo{author}{\bibfnamefont{R.}~\bibnamefont{Moshammer}},
  \bibinfo{author}{\bibfnamefont{A.}~\bibnamefont{Dorn}},
  \bibinfo{author}{\bibfnamefont{R.}~\bibnamefont{D\"orner}},
  \bibinfo{author}{\bibfnamefont{L.~P.~H.} \bibnamefont{Schmidt}},
  \bibnamefont{and}
  \bibinfo{author}{\bibfnamefont{H.}~\bibnamefont{Schmidt-B\"ocking}},
  \bibinfo{journal}{Rep. Prog. Phys.} \textbf{\bibinfo{volume}{66}},
  \bibinfo{pages}{1463} (\bibinfo{year}{2003}).

\bibitem[{\citenamefont{Eckle et~al.}(2008{\natexlab{a}})\citenamefont{Eckle,
  Smolarski, Schlup, Biegert, Staudte, Sch\"offler, Muller, D\"orner, and
  Keller}}]{Eckle_2008a}
\bibinfo{author}{\bibfnamefont{P.}~\bibnamefont{Eckle}},
  \bibinfo{author}{\bibfnamefont{M.}~\bibnamefont{Smolarski}},
  \bibinfo{author}{\bibfnamefont{F.}~\bibnamefont{Schlup}},
  \bibinfo{author}{\bibfnamefont{J.}~\bibnamefont{Biegert}},
  \bibinfo{author}{\bibfnamefont{A.}~\bibnamefont{Staudte}},
  \bibinfo{author}{\bibfnamefont{M.}~\bibnamefont{Sch\"offler}},
  \bibinfo{author}{\bibfnamefont{H.~G.} \bibnamefont{Muller}},
  \bibinfo{author}{\bibfnamefont{R.}~\bibnamefont{D\"orner}}, \bibnamefont{and}
  \bibinfo{author}{\bibfnamefont{U.}~\bibnamefont{Keller}},
  \bibinfo{journal}{Nature Phys.} \textbf{\bibinfo{volume}{4}},
  \bibinfo{pages}{565} (\bibinfo{year}{2008}{\natexlab{a}}).

\bibitem[{\citenamefont{Eckle et~al.}(2008{\natexlab{b}})\citenamefont{Eckle,
  Pfeiffer, Cirelli, Staudte, D\"orner, Muller, B\"uttiker, and
  Keller}}]{Eckle_2008b}
\bibinfo{author}{\bibfnamefont{P.}~\bibnamefont{Eckle}},
  \bibinfo{author}{\bibfnamefont{A.~N.} \bibnamefont{Pfeiffer}},
  \bibinfo{author}{\bibfnamefont{C.}~\bibnamefont{Cirelli}},
  \bibinfo{author}{\bibfnamefont{A.}~\bibnamefont{Staudte}},
  \bibinfo{author}{\bibfnamefont{R.}~\bibnamefont{D\"orner}},
  \bibinfo{author}{\bibfnamefont{H.~G.} \bibnamefont{Muller}},
  \bibinfo{author}{\bibfnamefont{M.}~\bibnamefont{B\"uttiker}},
  \bibnamefont{and} \bibinfo{author}{\bibfnamefont{U.}~\bibnamefont{Keller}},
  \bibinfo{journal}{Science} \textbf{\bibinfo{volume}{322}},
  \bibinfo{pages}{1525} (\bibinfo{year}{2008}{\natexlab{b}}).

\bibitem[{\citenamefont{Pfeiffer et~al.}(2012)\citenamefont{Pfeiffer, Cirelli,
  Smolarski, Dimitrovski, Abu-samha, Madsen, and Keller}}]{Pfeiffer_2012}
\bibinfo{author}{\bibfnamefont{A.~N.} \bibnamefont{Pfeiffer}},
  \bibinfo{author}{\bibfnamefont{C.}~\bibnamefont{Cirelli}},
  \bibinfo{author}{\bibfnamefont{M.}~\bibnamefont{Smolarski}},
  \bibinfo{author}{\bibfnamefont{D.}~\bibnamefont{Dimitrovski}},
  \bibinfo{author}{\bibfnamefont{M.}~\bibnamefont{Abu-samha}},
  \bibinfo{author}{\bibfnamefont{L.~B.} \bibnamefont{Madsen}},
  \bibnamefont{and} \bibinfo{author}{\bibfnamefont{U.}~\bibnamefont{Keller}},
  \bibinfo{journal}{Nature Phys.} \textbf{\bibinfo{volume}{8}},
  \bibinfo{pages}{76} (\bibinfo{year}{2012}).

\bibitem[{\citenamefont{Landsman et~al.}(2014)\citenamefont{Landsman, Weger,
  Maurer, Boge, Ludwig, Heuser, Cirelli, Gallmann, and
  Keller}}]{Landsman_2014o}
\bibinfo{author}{\bibfnamefont{A.~S.} \bibnamefont{Landsman}},
  \bibinfo{author}{\bibfnamefont{M.}~\bibnamefont{Weger}},
  \bibinfo{author}{\bibfnamefont{J.}~\bibnamefont{Maurer}},
  \bibinfo{author}{\bibfnamefont{R.}~\bibnamefont{Boge}},
  \bibinfo{author}{\bibfnamefont{A.}~\bibnamefont{Ludwig}},
  \bibinfo{author}{\bibfnamefont{S.}~\bibnamefont{Heuser}},
  \bibinfo{author}{\bibfnamefont{C.}~\bibnamefont{Cirelli}},
  \bibinfo{author}{\bibfnamefont{L.}~\bibnamefont{Gallmann}}, \bibnamefont{and}
  \bibinfo{author}{\bibfnamefont{U.}~\bibnamefont{Keller}},
  \bibinfo{journal}{Optica} \textbf{\bibinfo{volume}{1}}, \bibinfo{pages}{343}
  (\bibinfo{year}{2014}).

\bibitem[{\citenamefont{Landsman and Keller}(2014)}]{Landsman_2014b}
\bibinfo{author}{\bibfnamefont{A.~S.} \bibnamefont{Landsman}} \bibnamefont{and}
  \bibinfo{author}{\bibfnamefont{U.}~\bibnamefont{Keller}},
  \bibinfo{journal}{J. Phys. B} \textbf{\bibinfo{volume}{47}},
  \bibinfo{pages}{204024} (\bibinfo{year}{2014}).

\bibitem[{\citenamefont{Camus et~al.}(2017)\citenamefont{Camus, Yakaboylu,
  Fechner, Klaiber, Laux, Mi, Hatsagortsyan, Pfeifer, Keitel, and
  Moshammer}}]{Camus_2017}
\bibinfo{author}{\bibfnamefont{N.}~\bibnamefont{Camus}},
  \bibinfo{author}{\bibfnamefont{E.}~\bibnamefont{Yakaboylu}},
  \bibinfo{author}{\bibfnamefont{L.}~\bibnamefont{Fechner}},
  \bibinfo{author}{\bibfnamefont{M.}~\bibnamefont{Klaiber}},
  \bibinfo{author}{\bibfnamefont{M.}~\bibnamefont{Laux}},
  \bibinfo{author}{\bibfnamefont{Y.}~\bibnamefont{Mi}},
  \bibinfo{author}{\bibfnamefont{K.~Z.} \bibnamefont{Hatsagortsyan}},
  \bibinfo{author}{\bibfnamefont{T.}~\bibnamefont{Pfeifer}},
  \bibinfo{author}{\bibfnamefont{C.~H.} \bibnamefont{Keitel}},
  \bibnamefont{and}
  \bibinfo{author}{\bibfnamefont{R.}~\bibnamefont{Moshammer}},
  \bibinfo{journal}{Phys. Rev. Lett.} \textbf{\bibinfo{volume}{119}},
  \bibinfo{pages}{023201} (\bibinfo{year}{2017}).

\bibitem[{\citenamefont{Huismans et~al.}(2011)\citenamefont{Huismans,
  Rouz{\'{e}}e, Gijsbertsen, Jungmann, Smolkowska, Logman, L{\'{e}}pine,
  Cauchy, Zamith, Marchenko et~al.}}]{Huismans_2011}
\bibinfo{author}{\bibfnamefont{Y.}~\bibnamefont{Huismans}},
  \bibinfo{author}{\bibfnamefont{A.}~\bibnamefont{Rouz{\'{e}}e}},
  \bibinfo{author}{\bibfnamefont{A.}~\bibnamefont{Gijsbertsen}},
  \bibinfo{author}{\bibfnamefont{J.~H.} \bibnamefont{Jungmann}},
  \bibinfo{author}{\bibfnamefont{A.~S.} \bibnamefont{Smolkowska}},
  \bibinfo{author}{\bibfnamefont{P.~S. W.~M.} \bibnamefont{Logman}},
  \bibinfo{author}{\bibfnamefont{F.}~\bibnamefont{L{\'{e}}pine}},
  \bibinfo{author}{\bibfnamefont{C.}~\bibnamefont{Cauchy}},
  \bibinfo{author}{\bibfnamefont{S.}~\bibnamefont{Zamith}},
  \bibinfo{author}{\bibfnamefont{T.}~\bibnamefont{Marchenko}},
  \bibnamefont{et~al.}, \bibinfo{journal}{Science (New York, N.Y.)}
  \textbf{\bibinfo{volume}{331}}, \bibinfo{pages}{61} (\bibinfo{year}{2011}).

\bibitem[{\citenamefont{Bian et~al.}(2011)\citenamefont{Bian, Huismans,
  Smirnova, Yuan, Vrakking, and Bandrauk}}]{Bian_2011}
\bibinfo{author}{\bibfnamefont{X.-B.} \bibnamefont{Bian}},
  \bibinfo{author}{\bibfnamefont{Y.}~\bibnamefont{Huismans}},
  \bibinfo{author}{\bibfnamefont{O.}~\bibnamefont{Smirnova}},
  \bibinfo{author}{\bibfnamefont{K.-J.} \bibnamefont{Yuan}},
  \bibinfo{author}{\bibfnamefont{M.~J.~J.} \bibnamefont{Vrakking}},
  \bibnamefont{and} \bibinfo{author}{\bibfnamefont{A.~D.}
  \bibnamefont{Bandrauk}}, \bibinfo{journal}{Phys. Rev. A}
  \textbf{\bibinfo{volume}{84}}, \bibinfo{pages}{043420}
  (\bibinfo{year}{2011}).

\bibitem[{\citenamefont{Marchenko et~al.}(2011)\citenamefont{Marchenko,
  Huismans, Schafer, and Vrakking}}]{Marchenko_2011}
\bibinfo{author}{\bibfnamefont{T.}~\bibnamefont{Marchenko}},
  \bibinfo{author}{\bibfnamefont{Y.}~\bibnamefont{Huismans}},
  \bibinfo{author}{\bibfnamefont{K.~J.} \bibnamefont{Schafer}},
  \bibnamefont{and} \bibinfo{author}{\bibfnamefont{M.~J.~J.}
  \bibnamefont{Vrakking}}, \bibinfo{journal}{Phys. Rev. A}
  \textbf{\bibinfo{volume}{84}}, \bibinfo{pages}{053427}
  (\bibinfo{year}{2011}).

\bibitem[{\citenamefont{Huismans et~al.}(2012)\citenamefont{Huismans,
  Gijsbertsen, Smolkowska, Jungmann, Rouz{\'{e}}e, Logman, L{\'{e}}pine,
  Cauchy, Zamith, Marchenko et~al.}}]{Huismans_2012}
\bibinfo{author}{\bibfnamefont{Y.}~\bibnamefont{Huismans}},
  \bibinfo{author}{\bibfnamefont{A.}~\bibnamefont{Gijsbertsen}},
  \bibinfo{author}{\bibfnamefont{A.~S.} \bibnamefont{Smolkowska}},
  \bibinfo{author}{\bibfnamefont{J.~H.} \bibnamefont{Jungmann}},
  \bibinfo{author}{\bibfnamefont{A.}~\bibnamefont{Rouz{\'{e}}e}},
  \bibinfo{author}{\bibfnamefont{P.~S. W.~M.} \bibnamefont{Logman}},
  \bibinfo{author}{\bibfnamefont{F.}~\bibnamefont{L{\'{e}}pine}},
  \bibinfo{author}{\bibfnamefont{C.}~\bibnamefont{Cauchy}},
  \bibinfo{author}{\bibfnamefont{S.}~\bibnamefont{Zamith}},
  \bibinfo{author}{\bibfnamefont{T.}~\bibnamefont{Marchenko}},
  \bibnamefont{et~al.}, \bibinfo{journal}{Phys. Rev. Lett.}
  \textbf{\bibinfo{volume}{109}}, \bibinfo{pages}{013002}
  (\bibinfo{year}{2012}).

\bibitem[{\citenamefont{Landauer and Martin}(1994)}]{Landauer_1994}
\bibinfo{author}{\bibfnamefont{R.}~\bibnamefont{Landauer}} \bibnamefont{and}
  \bibinfo{author}{\bibfnamefont{T.}~\bibnamefont{Martin}},
  \bibinfo{journal}{Rev. Mod. Phys.} \textbf{\bibinfo{volume}{66}},
  \bibinfo{pages}{217} (\bibinfo{year}{1994}).

\bibitem[{\citenamefont{Sokolovski}(2008)}]{Sokolovski_2008}
\bibinfo{author}{\bibfnamefont{D.}~\bibnamefont{Sokolovski}}, in
  \emph{\bibinfo{booktitle}{Time in Quantum Mechanics}}, edited by
  \bibinfo{editor}{\bibfnamefont{J.~G.} \bibnamefont{Muga}},
  \bibinfo{editor}{\bibfnamefont{R.}~\bibnamefont{Mayato}}, \bibnamefont{and}
  \bibinfo{editor}{\bibfnamefont{I.~L.} \bibnamefont{Egusquiza}}
  (\bibinfo{publisher}{Springer Berlin / Heidelberg}, \bibinfo{year}{2008}),
  vol. \bibinfo{volume}{734} of \emph{\bibinfo{series}{Lecture Notes in
  Physics}}, pp. \bibinfo{pages}{195--233}.

\bibitem[{\citenamefont{Zimmermann et~al.}(2016)\citenamefont{Zimmermann,
  Mishra, Doran, Gordon, and Landsman}}]{Landsman_2016}
\bibinfo{author}{\bibfnamefont{T.}~\bibnamefont{Zimmermann}},
  \bibinfo{author}{\bibfnamefont{S.}~\bibnamefont{Mishra}},
  \bibinfo{author}{\bibfnamefont{B.~R.} \bibnamefont{Doran}},
  \bibinfo{author}{\bibfnamefont{D.~F.} \bibnamefont{Gordon}},
  \bibnamefont{and} \bibinfo{author}{\bibfnamefont{A.~S.}
  \bibnamefont{Landsman}}, \bibinfo{journal}{Phys. Rev. Lett.}
  \textbf{\bibinfo{volume}{116}}, \bibinfo{pages}{233603}
  (\bibinfo{year}{2016}).

\bibitem[{\citenamefont{Torlina et~al.}(2015)\citenamefont{Torlina, Morales,
  Kaushal, Ivanov, Kheifets, Zielinski, Scrinzi, Muller, Sukiasyan, Ivanov
  et~al.}}]{Torlina_2015}
\bibinfo{author}{\bibfnamefont{L.}~\bibnamefont{Torlina}},
  \bibinfo{author}{\bibfnamefont{F.}~\bibnamefont{Morales}},
  \bibinfo{author}{\bibfnamefont{J.}~\bibnamefont{Kaushal}},
  \bibinfo{author}{\bibfnamefont{I.}~\bibnamefont{Ivanov}},
  \bibinfo{author}{\bibfnamefont{A.}~\bibnamefont{Kheifets}},
  \bibinfo{author}{\bibfnamefont{A.}~\bibnamefont{Zielinski}},
  \bibinfo{author}{\bibfnamefont{A.}~\bibnamefont{Scrinzi}},
  \bibinfo{author}{\bibfnamefont{H.~G.} \bibnamefont{Muller}},
  \bibinfo{author}{\bibfnamefont{S.}~\bibnamefont{Sukiasyan}},
  \bibinfo{author}{\bibfnamefont{M.}~\bibnamefont{Ivanov}},
  \bibnamefont{et~al.}, \bibinfo{journal}{Nat. Phys.}
  \textbf{\bibinfo{volume}{11}}, \bibinfo{pages}{503} (\bibinfo{year}{2015}).

\bibitem[{\citenamefont{Ni et~al.}(2016)\citenamefont{Ni, Saalmann, and
  Rost}}]{Ni_2016}
\bibinfo{author}{\bibfnamefont{H.}~\bibnamefont{Ni}},
  \bibinfo{author}{\bibfnamefont{U.}~\bibnamefont{Saalmann}}, \bibnamefont{and}
  \bibinfo{author}{\bibfnamefont{J.-M.} \bibnamefont{Rost}},
  \bibinfo{journal}{Phys. Rev. Lett.} \textbf{\bibinfo{volume}{117}},
  \bibinfo{pages}{023002} (\bibinfo{year}{2016}).

\bibitem[{\citenamefont{Wigner}(1955)}]{Wigner_1955}
\bibinfo{author}{\bibfnamefont{E.~P.} \bibnamefont{Wigner}},
  \bibinfo{journal}{Phys. Rev.} \textbf{\bibinfo{volume}{98}},
  \bibinfo{pages}{145} (\bibinfo{year}{1955}).

\bibitem[{\citenamefont{Yakaboylu et~al.}(2013)\citenamefont{Yakaboylu,
  Klaiber, Bauke, Hatsagortsyan, and Keitel}}]{Yakaboylu_2013}
\bibinfo{author}{\bibfnamefont{E.}~\bibnamefont{Yakaboylu}},
  \bibinfo{author}{\bibfnamefont{M.}~\bibnamefont{Klaiber}},
  \bibinfo{author}{\bibfnamefont{H.}~\bibnamefont{Bauke}},
  \bibinfo{author}{\bibfnamefont{K.~Z.} \bibnamefont{Hatsagortsyan}},
  \bibnamefont{and} \bibinfo{author}{\bibfnamefont{C.~H.}
  \bibnamefont{Keitel}}, \bibinfo{journal}{Phys. Rev. A}
  \textbf{\bibinfo{volume}{88}}, \bibinfo{pages}{063421}
  (\bibinfo{year}{2013}).

\bibitem[{\citenamefont{Yakaboylu et~al.}(2014)\citenamefont{Yakaboylu,
  Klaiber, and Hatsagortsyan}}]{Yakaboylu_2014b}
\bibinfo{author}{\bibfnamefont{E.}~\bibnamefont{Yakaboylu}},
  \bibinfo{author}{\bibfnamefont{M.}~\bibnamefont{Klaiber}}, \bibnamefont{and}
  \bibinfo{author}{\bibfnamefont{K.~Z.} \bibnamefont{Hatsagortsyan}},
  \bibinfo{journal}{Phys. Rev. A} \textbf{\bibinfo{volume}{90}},
  \bibinfo{pages}{012116} (\bibinfo{year}{2014}).

\bibitem[{\citenamefont{Kaushal
  et~al.}(2015{\natexlab{a}})\citenamefont{Kaushal, Morales, and
  Smirnova}}]{Kaushal_2015a}
\bibinfo{author}{\bibfnamefont{J.}~\bibnamefont{Kaushal}},
  \bibinfo{author}{\bibfnamefont{F.}~\bibnamefont{Morales}}, \bibnamefont{and}
  \bibinfo{author}{\bibfnamefont{O.}~\bibnamefont{Smirnova}},
  \bibinfo{journal}{Phys. Rev. A} \textbf{\bibinfo{volume}{92}},
  \bibinfo{pages}{063405} (\bibinfo{year}{2015}{\natexlab{a}}).

\bibitem[{\citenamefont{Kaushal
  et~al.}(2015{\natexlab{b}})\citenamefont{Kaushal, Morales, Torlina, Ivanov,
  and Smirnova}}]{Kaushal_2015b}
\bibinfo{author}{\bibfnamefont{J.}~\bibnamefont{Kaushal}},
  \bibinfo{author}{\bibfnamefont{F.}~\bibnamefont{Morales}},
  \bibinfo{author}{\bibfnamefont{L.}~\bibnamefont{Torlina}},
  \bibinfo{author}{\bibfnamefont{M.}~\bibnamefont{Ivanov}}, \bibnamefont{and}
  \bibinfo{author}{\bibfnamefont{O.}~\bibnamefont{Smirnova}},
  \bibinfo{journal}{J. Phys. B} \textbf{\bibinfo{volume}{48}},
  \bibinfo{pages}{234002} (\bibinfo{year}{2015}{\natexlab{b}}).

\bibitem[{\citenamefont{Yudin and Ivanov}(2001)}]{Yudin_2001b}
\bibinfo{author}{\bibfnamefont{G.~L.} \bibnamefont{Yudin}} \bibnamefont{and}
  \bibinfo{author}{\bibfnamefont{M.~Y.} \bibnamefont{Ivanov}},
  \bibinfo{journal}{Phys. Rev. A} \textbf{\bibinfo{volume}{64}},
  \bibinfo{pages}{013409} (\bibinfo{year}{2001}).

\bibitem[{\citenamefont{Barth and Smirnova}(2013)}]{Barth_2013a}
\bibinfo{author}{\bibfnamefont{I.}~\bibnamefont{Barth}} \bibnamefont{and}
  \bibinfo{author}{\bibfnamefont{O.}~\bibnamefont{Smirnova}},
  \bibinfo{journal}{Phys. Rev. A} \textbf{\bibinfo{volume}{87}},
  \bibinfo{pages}{013433} (\bibinfo{year}{2013}).

\bibitem[{\citenamefont{Klaiber et~al.}(2015)\citenamefont{Klaiber,
  Hatsagortsyan, and Keitel}}]{Klaiber_2015}
\bibinfo{author}{\bibfnamefont{M.}~\bibnamefont{Klaiber}},
  \bibinfo{author}{\bibfnamefont{K.~Z.} \bibnamefont{Hatsagortsyan}},
  \bibnamefont{and} \bibinfo{author}{\bibfnamefont{C.~H.}
  \bibnamefont{Keitel}}, \bibinfo{journal}{Phys. Rev. Lett.}
  \textbf{\bibinfo{volume}{114}}, \bibinfo{pages}{083001}
  (\bibinfo{year}{2015}).

\bibitem[{\citenamefont{Keldysh}(1964)}]{Keldysh_1965}
\bibinfo{author}{\bibfnamefont{L.~V.} \bibnamefont{Keldysh}},
  \bibinfo{journal}{Zh. Eksp. Teor. Fiz.} \textbf{\bibinfo{volume}{47}},
  \bibinfo{pages}{1945} (\bibinfo{year}{1964}).

\bibitem[{\citenamefont{Faisal}(1973)}]{Faisal_1973}
\bibinfo{author}{\bibfnamefont{F.~H.~M.} \bibnamefont{Faisal}},
  \bibinfo{journal}{J. Phys. B} \textbf{\bibinfo{volume}{6}},
  \bibinfo{pages}{L89} (\bibinfo{year}{1973}).

\bibitem[{\citenamefont{Reiss}(1980)}]{Reiss_1980}
\bibinfo{author}{\bibfnamefont{H.~R.} \bibnamefont{Reiss}},
  \bibinfo{journal}{Phys. Rev. A} \textbf{\bibinfo{volume}{22}},
  \bibinfo{pages}{1786} (\bibinfo{year}{1980}).

\bibitem[{\citenamefont{Popruzhenko et~al.}(2008)\citenamefont{Popruzhenko,
  Paulus, and Bauer}}]{Popruzhenko_2008a}
\bibinfo{author}{\bibfnamefont{S.~V.} \bibnamefont{Popruzhenko}},
  \bibinfo{author}{\bibfnamefont{G.~G.} \bibnamefont{Paulus}},
  \bibnamefont{and} \bibinfo{author}{\bibfnamefont{D.}~\bibnamefont{Bauer}},
  \bibinfo{journal}{Phys. Rev. A} \textbf{\bibinfo{volume}{77}},
  \bibinfo{pages}{053409} (\bibinfo{year}{2008}).

\bibitem[{\citenamefont{Popruzhenko and Bauer}(2008)}]{Popruzhenko_2008b}
\bibinfo{author}{\bibfnamefont{S.~V.} \bibnamefont{Popruzhenko}}
  \bibnamefont{and} \bibinfo{author}{\bibfnamefont{D.}~\bibnamefont{Bauer}},
  \bibinfo{journal}{J. Mod. Opt.} \textbf{\bibinfo{volume}{55}},
  \bibinfo{pages}{2573} (\bibinfo{year}{2008}).

\bibitem[{\citenamefont{Torlina and Smirnova}(2012)}]{Torlina_2012}
\bibinfo{author}{\bibfnamefont{L.}~\bibnamefont{Torlina}} \bibnamefont{and}
  \bibinfo{author}{\bibfnamefont{O.}~\bibnamefont{Smirnova}},
  \bibinfo{journal}{Phys. Rev. A} \textbf{\bibinfo{volume}{86}},
  \bibinfo{pages}{043408} (\bibinfo{year}{2012}).

\bibitem[{\citenamefont{Torlina et~al.}(2012)\citenamefont{Torlina, Ivanov,
  Walters, and Smirnova}}]{Torlina_2012b}
\bibinfo{author}{\bibfnamefont{L.}~\bibnamefont{Torlina}},
  \bibinfo{author}{\bibfnamefont{M.}~\bibnamefont{Ivanov}},
  \bibinfo{author}{\bibfnamefont{Z.~B.} \bibnamefont{Walters}},
  \bibnamefont{and} \bibinfo{author}{\bibfnamefont{O.}~\bibnamefont{Smirnova}},
  \bibinfo{journal}{Phys. Rev. A} \textbf{\bibinfo{volume}{86}},
  \bibinfo{pages}{043409} (\bibinfo{year}{2012}).

\bibitem[{\citenamefont{Torlina et~al.}(2013)\citenamefont{Torlina, Kaushal,
  and Smirnova}}]{Torlina_2013}
\bibinfo{author}{\bibfnamefont{L.}~\bibnamefont{Torlina}},
  \bibinfo{author}{\bibfnamefont{J.}~\bibnamefont{Kaushal}}, \bibnamefont{and}
  \bibinfo{author}{\bibfnamefont{O.}~\bibnamefont{Smirnova}},
  \bibinfo{journal}{Phys. Rev. A} \textbf{\bibinfo{volume}{88}},
  \bibinfo{pages}{053403} (\bibinfo{year}{2013}).

\bibitem[{\citenamefont{Becker et~al.}(1989)\citenamefont{Becker, McIver, and
  Confer}}]{Becker_1989}
\bibinfo{author}{\bibfnamefont{W.}~\bibnamefont{Becker}},
  \bibinfo{author}{\bibfnamefont{J.~K.} \bibnamefont{McIver}},
  \bibnamefont{and} \bibinfo{author}{\bibfnamefont{M.}~\bibnamefont{Confer}},
  \bibinfo{journal}{Phys. Rev. A} \textbf{\bibinfo{volume}{40}},
  \bibinfo{pages}{6904} (\bibinfo{year}{1989}).

\bibitem[{\citenamefont{Figueira~de Morisson~Faria
  et~al.}(2002)\citenamefont{Figueira~de Morisson~Faria, Schomerus, and
  Becker}}]{Faria_2002}
\bibinfo{author}{\bibfnamefont{C.}~\bibnamefont{Figueira~de Morisson~Faria}},
  \bibinfo{author}{\bibfnamefont{H.}~\bibnamefont{Schomerus}},
  \bibnamefont{and} \bibinfo{author}{\bibfnamefont{W.}~\bibnamefont{Becker}},
  \bibinfo{journal}{Phys. Rev. A} \textbf{\bibinfo{volume}{66}},
  \bibinfo{pages}{043413} (\bibinfo{year}{2002}).

\bibitem[{\citenamefont{Becker et~al.}(1990)\citenamefont{Becker, Long, and
  McIver}}]{Becker_1990a}
\bibinfo{author}{\bibfnamefont{W.}~\bibnamefont{Becker}},
  \bibinfo{author}{\bibfnamefont{S.}~\bibnamefont{Long}}, \bibnamefont{and}
  \bibinfo{author}{\bibfnamefont{J.~K.} \bibnamefont{McIver}},
  \bibinfo{journal}{Phys. Rev. A} \textbf{\bibinfo{volume}{41}},
  \bibinfo{pages}{4112} (\bibinfo{year}{1990}).

\bibitem[{\citenamefont{Becker et~al.}(1994)\citenamefont{Becker, Long, and
  McIver}}]{Becker_1994b}
\bibinfo{author}{\bibfnamefont{W.}~\bibnamefont{Becker}},
  \bibinfo{author}{\bibfnamefont{S.}~\bibnamefont{Long}}, \bibnamefont{and}
  \bibinfo{author}{\bibfnamefont{J.~K.} \bibnamefont{McIver}},
  \bibinfo{journal}{Phys. Rev. A} \textbf{\bibinfo{volume}{50}},
  \bibinfo{pages}{1540} (\bibinfo{year}{1994}).

\bibitem[{\citenamefont{Kopold et~al.}(2000)\citenamefont{Kopold, Becker,
  Rottke, and Sandner}}]{Kopold_2000}
\bibinfo{author}{\bibfnamefont{R.}~\bibnamefont{Kopold}},
  \bibinfo{author}{\bibfnamefont{W.}~\bibnamefont{Becker}},
  \bibinfo{author}{\bibfnamefont{H.}~\bibnamefont{Rottke}}, \bibnamefont{and}
  \bibinfo{author}{\bibfnamefont{W.}~\bibnamefont{Sandner}},
  \bibinfo{journal}{Phys. Rev. Lett.} \textbf{\bibinfo{volume}{85}},
  \bibinfo{pages}{3781} (\bibinfo{year}{2000}).

\bibitem[{\citenamefont{Becker et~al.}(2002)\citenamefont{Becker, Grasbon,
  Kopold, Milo\u{s}evi\'c, Paulus, and Walther}}]{Becker_2002}
\bibinfo{author}{\bibfnamefont{W.}~\bibnamefont{Becker}},
  \bibinfo{author}{\bibfnamefont{F.}~\bibnamefont{Grasbon}},
  \bibinfo{author}{\bibfnamefont{R.}~\bibnamefont{Kopold}},
  \bibinfo{author}{\bibfnamefont{D.}~\bibnamefont{Milo\u{s}evi\'c}},
  \bibinfo{author}{\bibfnamefont{G.~G.} \bibnamefont{Paulus}},
  \bibnamefont{and} \bibinfo{author}{\bibfnamefont{H.}~\bibnamefont{Walther}},
  \bibinfo{journal}{Adv. Atom. Mol. Opt. Phys.} \textbf{\bibinfo{volume}{48}},
  \bibinfo{pages}{35} (\bibinfo{year}{2002}).

\bibitem[{\citenamefont{\ifmmode \check{C}\else
  \v{C}\fi{}erki\ifmmode~\acute{c}\else \'{c}\fi{}
  et~al.}(2009)\citenamefont{\ifmmode \check{C}\else
  \v{C}\fi{}erki\ifmmode~\acute{c}\else \'{c}\fi{},
  Hasovi\ifmmode~\acute{c}\else \'{c}\fi{}, Milo\ifmmode \check{s}\else
  \v{s}\fi{}evi\ifmmode~\acute{c}\else \'{c}\fi{}, and Becker}}]{Cerkic_2009}
\bibinfo{author}{\bibfnamefont{A.}~\bibnamefont{\ifmmode \check{C}\else
  \v{C}\fi{}erki\ifmmode~\acute{c}\else \'{c}\fi{}}},
  \bibinfo{author}{\bibfnamefont{E.}~\bibnamefont{Hasovi\ifmmode~\acute{c}\else
  \'{c}\fi{}}}, \bibinfo{author}{\bibfnamefont{D.~B.} \bibnamefont{Milo\ifmmode
  \check{s}\else \v{s}\fi{}evi\ifmmode~\acute{c}\else \'{c}\fi{}}},
  \bibnamefont{and} \bibinfo{author}{\bibfnamefont{W.}~\bibnamefont{Becker}},
  \bibinfo{journal}{Phys. Rev. A} \textbf{\bibinfo{volume}{79}},
  \bibinfo{pages}{033413} (\bibinfo{year}{2009}).

\bibitem[{\citenamefont{Milo\ifmmode \check{s}\else
  \v{s}\fi{}evi\ifmmode~\acute{c}\else \'{c}\fi{}}(2014)}]{Milosevic_2014}
\bibinfo{author}{\bibfnamefont{D.~B.} \bibnamefont{Milo\ifmmode \check{s}\else
  \v{s}\fi{}evi\ifmmode~\acute{c}\else \'{c}\fi{}}}, \bibinfo{journal}{Phys.
  Rev. A} \textbf{\bibinfo{volume}{90}}, \bibinfo{pages}{063423}
  (\bibinfo{year}{2014}).

\bibitem[{\citenamefont{Krajewska et~al.}(2010)\citenamefont{Krajewska,
  Kami\'nski, and W\'odkiewicz}}]{Krajewska_2010}
\bibinfo{author}{\bibfnamefont{K.}~\bibnamefont{Krajewska}},
  \bibinfo{author}{\bibfnamefont{J.}~\bibnamefont{Kami\'nski}},
  \bibnamefont{and}
  \bibinfo{author}{\bibfnamefont{K.}~\bibnamefont{W\'odkiewicz}},
  \bibinfo{journal}{Opt. Commun.} \textbf{\bibinfo{volume}{283}},
  \bibinfo{pages}{843 } (\bibinfo{year}{2010}).

\bibitem[{Sup()}]{Suppl_material}
\bibinfo{howpublished}{See the Supplemental Materials for the details.}

\bibitem[{\citenamefont{Lein}(2011)}]{Lein_2011}
\bibinfo{author}{\bibfnamefont{M.}~\bibnamefont{Lein}}, \bibinfo{journal}{J.
  Mod. Opt.} \textbf{\bibinfo{volume}{58}}, \bibinfo{pages}{1188}
  (\bibinfo{year}{2011}).

\bibitem[{\citenamefont{Gribakin and Kuchiev}(1997)}]{Gribakin_1997}
\bibinfo{author}{\bibfnamefont{G.}~\bibnamefont{Gribakin}} \bibnamefont{and}
  \bibinfo{author}{\bibfnamefont{M.}~\bibnamefont{Kuchiev}},
  \bibinfo{journal}{Phys. Rev. A} \textbf{\bibinfo{volume}{55}},
  \bibinfo{pages}{3760} (\bibinfo{year}{1997}).

\bibitem[{\citenamefont{Delone and Krainov}(1991)}]{Delone_1991}
\bibinfo{author}{\bibfnamefont{N.~B.} \bibnamefont{Delone}} \bibnamefont{and}
  \bibinfo{author}{\bibfnamefont{V.~P.} \bibnamefont{Krainov}},
  \bibinfo{journal}{J. Opt. Soc. Am. B} \textbf{\bibinfo{volume}{8}},
  \bibinfo{pages}{1207} (\bibinfo{year}{1991}).

\bibitem[{\citenamefont{Ivanov et~al.}(1996)\citenamefont{Ivanov, Brabec, and
  Burnett}}]{Ivanov_1996}
\bibinfo{author}{\bibfnamefont{M.~Y.} \bibnamefont{Ivanov}},
  \bibinfo{author}{\bibfnamefont{T.}~\bibnamefont{Brabec}}, \bibnamefont{and}
  \bibinfo{author}{\bibfnamefont{N.}~\bibnamefont{Burnett}},
  \bibinfo{journal}{Phys. Rev. A} \textbf{\bibinfo{volume}{54}},
  \bibinfo{pages}{742} (\bibinfo{year}{1996}).

\end{thebibliography}

\end{document}